%% file: globulars.tex
\documentclass[fleqn,usenatbib]{mnras}
\usepackage{newtxtext,newtxmath}
\usepackage{multirow}
\usepackage{hyperref}

\usepackage[T1]{fontenc}
\usepackage{ae,aecompl}

\usepackage{graphicx}	% Including figure files
\usepackage{amsmath}	% Advanced maths commands
\usepackage{amssymb}	% Extra maths symbols

\title[GC mass-to-light ratios at high metallicity]{The WAGGS project - III. Discrepant mass-to-light ratios of Galactic globular clusters at high metallicity}

\author[H. Dalgleish et al.]{H. Dalgleish$^{1}$\thanks{E-mail: h.s.dalgleish@2016.ljmu.ac.uk},
S. Kamann$^{1}$,
C. Usher$^{1}$, 
H. Baumgardt$^{2}$, 
N. Bastian$^{1}$, \newauthor
J. Veitch-Michaelis$^{1}$,
A. Bellini$^{3}$, 
S. Martocchia$^{4,1}$, 
G. S. Da Costa$^{5}$, \newauthor
D. Mackey$^{5}$,
S. Bellstedt$^{6}$
N. Pastorello$^{7}$,
P. Cerulo$^{8}$
\\
$^{1}$Astrophysics Research Institute, Liverpool John Moores University, Liverpool, L3 5RF, UK \\
$^{2}$School of Mathematics and Physics, The University of Queensland, St. Lucia, QLD 4072, Australia \\
$^{3}$Space Telescope Science Institute, 3800 San Martin Drive, Baltimore, MD 21218, USA \\
$^{4}$European Southern Observatory, Karl-Schwarzschild-Strasse 2, D-85748, Garching bei M{\"u}nchen \\
$^{5}$Research School of Astronomy and Astrophysics, Australian National University, Canberra, ACT 2611, Australia \\
$^{6}$ICRAR, The University of Western Australia, 7 Fairway, Crawley WA 6009, Australia \\
$^{7}$BlueScope, Level 11, 120 Collins Street, Melbourne VIC 3000, Australia \\
$^{8}$Department of Astronomy, Universidad de Concepci\'{o}n, Casilla 160-C, Concepci\'{o}n, Chile
}

\date{Accepted XXX. Received YYY; in original form ZZZ}

\pubyear{2018}

\begin{document}
\label{firstpage}
\pagerange{\pageref{firstpage}--\pageref{lastpage}}
\maketitle

\begin{abstract}
Observed mass-to-light ratios ($M/L$) of metal-rich globular clusters (GCs) disagree with theoretical predictions. This discrepancy is of fundamental importance since stellar population models provide the stellar masses that underpin most of extragalactic astronomy, near and far. We have derived radial velocities for 1,622 stars located in the centres of 59 Milky Way GCs --- twelve of which have no previous kinematic information --- using integral-field unit data from the WAGGS project. Using $N$-body models, we determine dynamical masses and $M/L_\mathrm{V}$ ratios for the studied clusters. Our sample includes NGC 6528 and NGC 6553, which extend the metallicity range of GCs with measured $M/L$ up to [Fe/H] $\sim -0.1$ dex. 
We find that metal-rich clusters have $M/L_\mathrm{V}$ more than 2 times lower than what is predicted by simple stellar population models. This confirms that the discrepant $M/L$--[Fe/H] relation remains a serious concern. We explore how our findings relate to previous observations, and the potential causes for the divergence, which we conclude is most likely due to dynamical effects.
\end{abstract}


%%%%%%%%%%%%%%%%%%%%%%%%%%%%%%%%%%%%%%%%%%%%%%%%%%

%%%%%%%%%%%%%%%%% BODY OF PAPER %%%%%%%%%%%%%%%%%%

\section{Introduction}

Globular clusters (GCs) contain large numbers ($10^5-10^6$) of stars of roughly the same age and metallicity. For this reason, GCs are ideal laboratories for the study of the formation and evolution of stars and their host galaxies (e.g. \citealt{1998gcs..book.....A}). 
In particular, internal cluster kinematics have a key role in our understanding of GCs and their origins. 

When research on this topic began in the late 1970s, studies depended on limited radial velocity information to derive velocity dispersions, dynamical masses, and other important cluster parameters  (\citealt{1993ASPC...50..357P}, and references therein).
%Quantities like the mass-to-light ($M/L$) ratio provide an insight to the associated initial mass functions (IMF) of GCs as well as stringent constraints on simple stellar population (SSP) models. A thorough understanding of the $M/L$ ratios is particularly crucial for the mass determination of external galaxies, for which the IMF is a considerable source of uncertainty in stellar population studies \citep{2013ARA&A..51..393C}.
%Dynamical \textit{M}/\textit{L} ratios depend on accurate velocity dispersions needed for mass derivations. 
%This work was considerably expanded by \citet{2005ApJS..161..304M} who presented dynamical \textit{M}/\textit{L} ratios for 57 star clusters. 
King (\citeyear{1966AJ.....71...64K}) models --- and variations thereof --- have commonly been used for this work, however static models such as these suffer limitations; they cannot account for cluster relaxation, a result of gravitational interactions of their member stars. Therefore, phenomena such as mass segregation, mass loss, or core collapse cannot be understood using these one-component static models. Multi-component static models, on the other hand, can take such evolutionary dynamical processes into account (see e.g. \citealt{2019arXiv190913093T}). They can be analytical, like the isotropic multimass King-Michie models \citep{1979AJ.....84..752G}, \texttt{LIMEPY} models \citep{2015MNRAS.454..576G}, \textit{N}-body models \citep{2011MNRAS.411.1989Z,2014MNRAS.445.3435H,2016MNRAS.458.1450W,2017MNRAS.464.2174B,2018MNRAS.478.1520B}, or Monte-Carlo models. Most recently, \citet{2019MNRAS.482.5138B} have used \textit{N}-body models to determine masses, structural parameters, and mass-to-light ratios of 144 GCs, where they also included \textit{Gaia} DR2 proper motions besides radial velocities.
% take into account mass segregation and core collapse. Dynamic and evolutionary methods such as isotropic multimass King-Michie models \citep{1979AJ.....84..752G}, \texttt{LIMEPY} models \citep{2015MNRAS.454..576G}, or \textit{N}-body models \citep{2011MNRAS.411.1989Z,2014MNRAS.445.3435H,2016MNRAS.458.1450W,2017MNRAS.464.2174B,2018MNRAS.478.1520B} can better describe GCs by taking into account dynamical mass segregation and the distribution of low-mass stars that cannot easily be observed. Most recently, \citet{2019MNRAS.482.5138B} have determined masses, structural parameters, and mass-to-light ratios of 144 GCs, where they also include \textit{Gaia} DR2 proper motions. 

The available kinematic data are often a significant limitation when constraining cluster parameters via evolutionary models. \textit{Gaia} and most ground-based multi-object spectrographs like VLT/FLAMES or Keck/DEIMOS are unable to observe a large number of stars within the core radii of GCs because of the strong stellar crowding in those regions. Given the short relaxation times and expected overdensities of stellar remnants near the centres (due to mass segregation), it is likely a key area of parameter space has been missed. Now with $HST$ and the development of integral-field units (IFUs) like MUSE, studies can measure the motions of thousands of stars --- including those in the cluster centres --- for the first time (e.g. \citealt{2014ApJ...797..115B,2017ApJ...844..167B,2018MNRAS.473.5591K,2018ApJ...860...50F}). This has opened up new avenues to uncover the populations of central stellar remnants, like stellar-mass black holes (e.g. \citealt{2018MNRAS.475L..15G,2019A&A...632A...3G,2019MNRAS.488.5340B}), which can further aid our understanding of the internal structures of GCs.

One puzzle yet to be solved is the notable discrepancy between theoretical predictions and observations of GC $M/L$ ratios. Simple stellar population (SSP) models predict that the mass-to-light ratio in the V band ($M/L_\text{V}$) should increase with metallicity --- given a constant initial mass function (IMF) --- as a result of line blanketing (e.g. \citealt{2003MNRAS.344.1000B,2005MNRAS.362..799M,2010ApJ...712..833C}). However, \citet{1997ApJ...474L..19D} and \citet{2009AJ....138..547S,2011AJ....142....8S} found that for the globular clusters in M31, $M/L$ decreases with [Fe/H].
In the Milky Way, the situation appears to be similar; \citet{2015AJ....149...53K} showed that $M/L$ is $\gtrsim2$ times lower than expected for clusters at the metal-rich end. %\textbf{[Gieles: the variation of the M/L is due to poor mass modelling].}
Equally, \citet{2017MNRAS.464.2174B} found that the observed $M/L$-[Fe/H] relation also disagrees with SSP models: they find no change of $M/L$ with cluster metallicity. Very few metal-rich Galactic clusters ([Fe/H] $> -0.5$ dex) were included in the studies of \citet{2015AJ....149...53K} and \citet{2017MNRAS.464.2174B}, however, so it is unclear if the discrepancy for MW clusters is as pronounced as it is for M31 towards solar metallicity.
This is further emphasised by \citet{2015MNRAS.448L..94S}, who discuss the challenges of reliably measuring the $M/L$ of star clusters from integrated light.

Understanding where the discrepancy between observed $M/L$ and SSP model predictions originates is key. SSP models are used to determine a wide range of important properties --- including stellar masses, star formation histories and metallicities --- from the integrated light of galaxies and extragalactic star clusters (e.g. \citealt{2009ApJS..185..253G,2013ARA&A..51..393C,2019MNRAS.484.2388M}).
Furthermore, the mismatch between observed stellar $M/L$ and SSP model predictions has been used to constrain the IMF \citep[e.g.][]{2012Natur.484..485C,2013ApJ...765...25N}. Hence, more high-resolution observations of metal-rich clusters are needed to confirm the discrepancy between theory and observation.

%"Such correlations should in principle exist since for example the loss of low-mass stars decreases the M/L ratio of a globular cluster (Baumgardt & Makino 2003; Kruijssen & Mieske 2009). In addition, the loss of stellar remnants can decrease the M/L ratio of a globular cluster (Bianchini et al. 2017). The reason for the absence of a correlation could be that the resulting change in M/L ratio is too small to cause a noticeable difference or is compensated for by the correlation between mass function slope and cluster metallicity."

WAGGS, the WiFeS Atlas of Galactic Globular cluster Spectra survey \citep{2017MNRAS.468.3828U,2019MNRAS.482.1275U} has already significantly extended GC observations to include younger and more metal-rich clusters. The survey is also particularly advantageous since it covers the central regions of all clusters ($\lesssim 20$ arcsec).
Our study uses the WAGGS survey to fill this gap in the literature, with newly-determined central velocity dispersions, dynamical masses, and $M/L_\mathrm{V}$ for 59 globular clusters in the Milky Way. For twelve of these GCs, $M/L_\mathrm{V}$ have never before been derived, due to the lack of kinematic information.

We organise the paper as follows: our observations are described in Section 2, followed by a description of our data reduction and analysis (Section 3). The results and discussion are presented in Sections 4 and 5 respectively, with concluding remarks made in Section 6.
%The latest spectroscopic surveys have provided a good picture of velocity dispersion profiles and masses of the majority of globular clusters in the Milky Way, however this picture is incomplete. Technological limitations have prevented a view on the central, and therefore densest, regions of GCs, leaving central velocity dispersions of most clusters to be estimated. 

\begin{figure}
    \centering
    \includegraphics[width=0.47\textwidth]{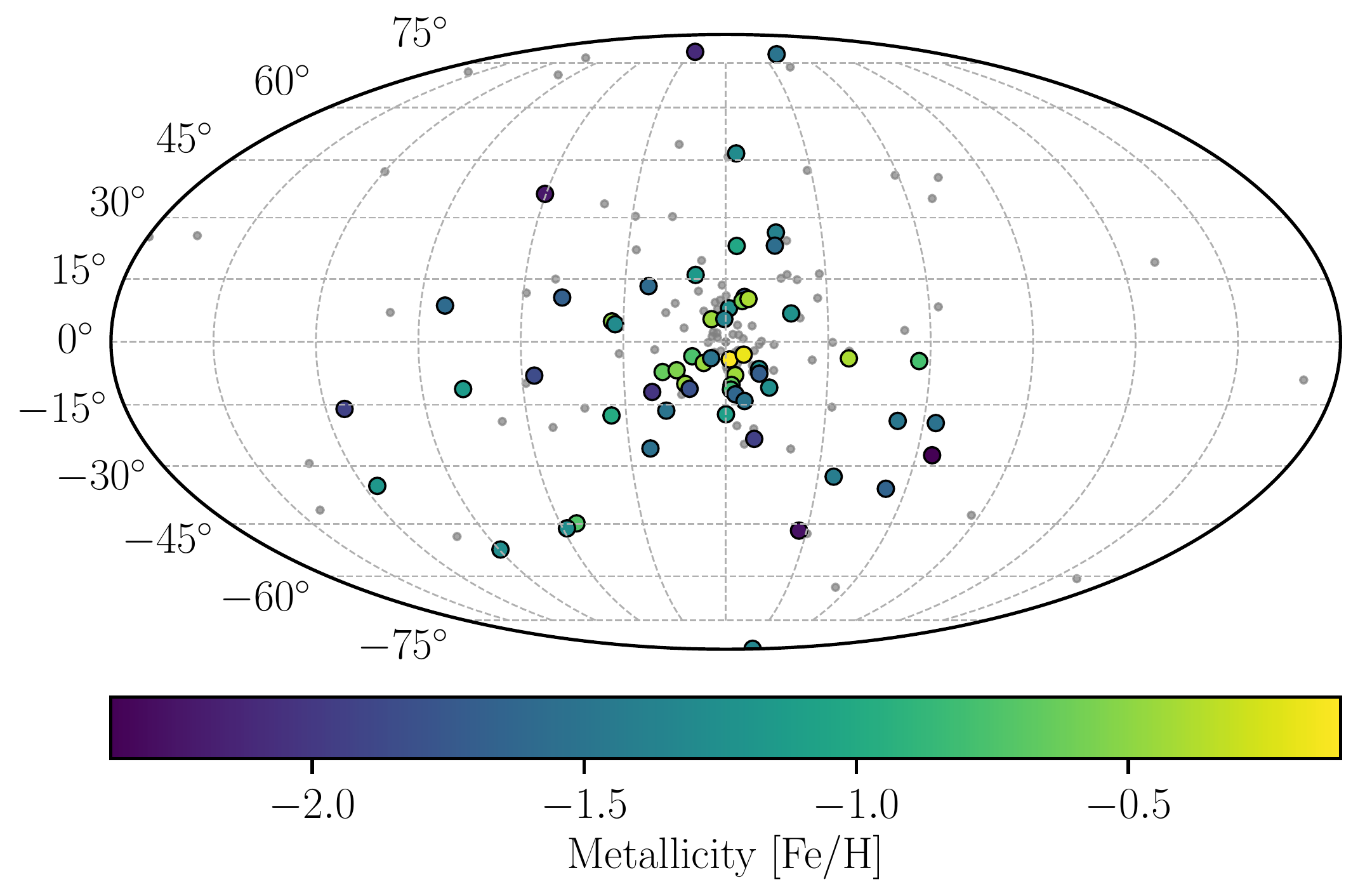}
    \caption{Milky Way globular clusters projected in Mollweide space in Galactic coordinates. The subset of 59 globular clusters from the WAGGS survey used in this study are highlighted: the most metal-rich clusters appear yellow and are found towards the bulge, and the most metal-poor clusters are in purple, typically found towards the halo. Any remaining Milky Way GCs are shown as grey dots \citep{1996AJ....112.1487H,2010arXiv1012.3224H}.}
    \label{fig:mollweide}
\end{figure}

\section{Observations}

For this work, we studied a subset of 59 GCs taken from the WAGGS survey (Figure \ref{fig:mollweide}). The clusters were selected on the basis that they either have $HST$ photometry and imaging publicly available, they are metal-rich (i.e. NGCs 6528 and 6553), or there are fewer than 25 radial velocity measurements in the \citet{2019MNRAS.482.5138B} sample. %(i.e. NGCs 5946, 6325, 6333, 6342, 6355, 6356, 6380, 6453, 6517, 6642, 6760, and 7006).

WAGGS observations were made between 2015 and 2018 using WiFeS, the Wide-Field Spectrograph, on the 2.3-m ANU telescope \citep{2007Ap&SS.310..255D,2010Ap&SS.327..245D}.
WiFeS is an IFU spectrograph with a field-of-view of $38 \times 25$ arcsec.
We used the four higher resolution gratings (U7000, B7000, R7000, and I7000) in two setups to cover the wavelength range $3300-9000$ \AA{} with spectral resolution, R $\sim 6800$. Further details regarding the observations and data reduction can be found in \citet{2017MNRAS.468.3828U}.
In addition to the observations described in \citet{2017MNRAS.468.3828U,2019MNRAS.482.1275U}, we include new observations of NGCs 6325, 6355, 6380, 6453, 6528, 6553, and 6760 carried out between September 10th--13th 2018. Of these, NGC 6528 and NGC 6553 had already been observed before.
Unlike most WAGGS observations, where only a single pointing centred on the GC centre was observed, for the new NGC 6528 and NGC 6553 observations we tiled the centres of these GCs with four pointings.
%We also utilised previously observed multiple pointings of NGC 104, 288, 3201, 6121, 6397 and 6496.
We reduced the September 2018 observations using PyWiFeS \citep{2014Ap&SS.349..617C, 2014ascl.soft02034C} in the same manner as described in \citet{2017MNRAS.468.3828U}.
%The pixel scale is 1 $\times$ 1 ''. The data were obtained for the  which is still ongoing. Currently the survey comprises central regions of 84 Milky Way GCs and 30 of the galaxy's satellites (Fornax, LMC, and SMC).

\input{table_hst_obvs.tex}

\subsection{Photometry}
\label{sec:photometry}

Photometry for the clusters is required to extract the spectra from the WAGGS datacubes (see Section \ref{sec:pampelmuse}) and comes from previous literature wherever possible. For the majority of GCs, we obtained existing $HST$ photometry from the UV Legacy Survey \citep{2017AJ....153...19S}. NGC 6528 photometry was sourced from the Hubble Legacy Archive, in programs 5436 (PI: Ortolani), 8696 (PI: Feltzing), 9453 (PI: Brown), 11664 (PI: Brown), and the photometry for NGC 6356 and NGC 7006 were obtained from  \citet{2002A&A...391..945P} and \citet{2011ApJ...738...74D}, respectively.
For eight of the clusters, photometry was not available (i.e. NGCs 5946, 6325, 6342, 6355, 6380, 6453, 6517, and 6553). These remaining clusters were reduced using \textit{HST} observations as shown in Table~\ref{tabHSTobs}. 

The reduction of all \textit{HST} exposures is based on \texttt{\_flc} images. These images have been corrected for the flat-field and charge-transfer efficiency effects \citep{2010PASP..122.1035A,2018acs..rept....4R}
%\footnote{CTE (charge-transfer efficiency) defects are caused by traps in the silicon lattice of the CCD. During image readout, electrons captured by these traps are released at a later stage and are not shuffled down toward the readout register together with their original pixel, but are released at a later stage. This causes point-like sources to appear as ``water drops''; the effect becomes larger the farther the traps are from the readout registers and the fainter the stars (see \citealt{2010PASP..122.1035A,2018acs..rept....4R} for more details).} 
and bias-subtracted via the standard \textit{HST} pipelines, and contain the unresampled pixel data for stellar-profile fitting. 

\subsubsection{NGCs 5946, 6325, 6342, 6355, 6380, 6453, 6517}

The data for the first seven clusters are obtained from GO-11628 (P.I.:\ E.\ Noyola) and we measured the stellar photometry as follows. First, for each exposure, we derived state-of-the-art, spatially variable PSF models by perturbing the library PSF models created by Jay Anderson\footnote{\url{www.stsci.edu/~jayander/WFC3/}} in order to account for telescope breathing effects (\citealt{2008acs..rept....3D}, see also \citealt{2017ApJ...842....6B,2018acs..rept....8B}). Using these PSFs, we measure stellar positions and fluxes of bright sources in each exposure (including saturated stars) using the \texttt{FORTRAN} code \texttt{HST1pass} (Anderson in prep., see \citealt{2018ApJ...853...86B} for details). Photometry of the saturated stars is obtained by including all the relevant flux from the bled-into pixels following the prescriptions given in \cite{2010wfc..rept...10G}. This technique allows us to achieve 5\% photometric RMS three magnitudes above saturation. Stellar positions and magnitudes in these single-exposure catalogues are corrected for geometric distortion and pixel-area \citep{2011PASP..123..622B}.

We used the \textit{Gaia} DR2 catalogue (\citealt{2016A&A...595A...1G,2018A&A...616A...1G}) to define a reference frame with a pixel scale of exactly 50 mas. We transformed stellar positions from each single-exposure catalogue on to the reference frame by means of six-parameter linear transformations using bright, unsaturated stars.

Next, we obtain best estimates of positions and fluxes for all possible sources, using the \texttt{FORTRAN} software package \texttt{KS2} (Anderson, in prep., see \citealt{2017ApJ...842....6B} for details). \texttt{KS2} takes the image-tailored PSF models and six-parameter linear transformations of each catalogue on to the reference frame, and uses all the exposures simultaneously to find, measure and subtract sources through different finding waves. \texttt{KS2} measures stellar photometry using three different methods, each of which is best suited for specific brightness regimes. Since our primary objective is to derive reliable mass-to-light ratios, we relied on \texttt{KS2}'s ``method 1'', which offers the best photometry for intermediate and high brightness regimes. 

%Instrumental magnitudes were converted to the VEGAMAG flight system using the prescriptions given in \cite{2017ApJ...842....6B} and the zero-points published on the WFC3 webpage\footnote{\url{www.stsci.edu/HST/instrumentation/wfc3/data-analysis/photometric-calibration/uvis-photometric-calibration}}.

\subsubsection{NGC 6553}

The NGC 6553 photometric catalogue is obtained from GO-10573 (P.I.: M. Mateo). We corrected for pixel-area effects by applying Pixel Area Maps to each \texttt{\_flc} image as well as cosmic ray contamination by using the L.A. Cosmic algorithm \citep{2001PASP..113.1420V}.
We followed the same strategy as \citet{2014ApJ...791L...4D} for the photometric analysis: 

First we used DAOPHOTIV \citep{1987PASP...99..191S} independently on each chip and filter. Several hundreds of bright and isolated stars were selected in order to model the PSF. All available analytic functions were considered for the PSF fitting (i.e. Gauss, Moffat, Lorentz and Penny functions), leaving the PSF free to spatially vary to the first-order. For each image, we fit all the star-like sources with the obtained PSF using a threshold of 3$\sigma$ from the local background. Next, we used ALLFRAME \citep{1984PASP...96..128S} to generate a star list with stars detected in at least three out of the five images, when putting all the filters together, one chip at a time. The final star lists for each image and chip were cross-correlated using DAOMATCH, followed by magnitude means and standard deviations with DAOMASTER. We obtained the final catalogue through matching the star lists for each filter via DAOMATCH and DAOMASTER. Finally, instrumental magnitudes were converted to the VEGAMAG photometric system given the prescriptions and zero-points reported on the ACS webpage\footnote{\url{http://acszeropoints.stsci.edu/}} and in \citet{2005PASP..117.1049S}.

\section{Data Reduction and Analysis}

\subsection{Stellar spectra}
\label{sec:pampelmuse}

In order to extract the stellar spectra from the WAGGS data cubes, we used \textsc{PampelMuse} \citep{2013A&A...549A..71K}. The software optimally extracts the spectrum of each resolved star from a cube using an analytical model of the point-spread function (PSF). The model parameters can be wavelength-dependent and are optimized during the fit. The coordinates of each resolved source are inferred from the source catalogs described in Sect.~\ref{sec:photometry}. The PSF-fitting technique allows one to de-blend clean stellar spectra even in crowded stellar fields such as the centres of GCs.

\begin{figure*}
    \centering
    \includegraphics[width=\textwidth]{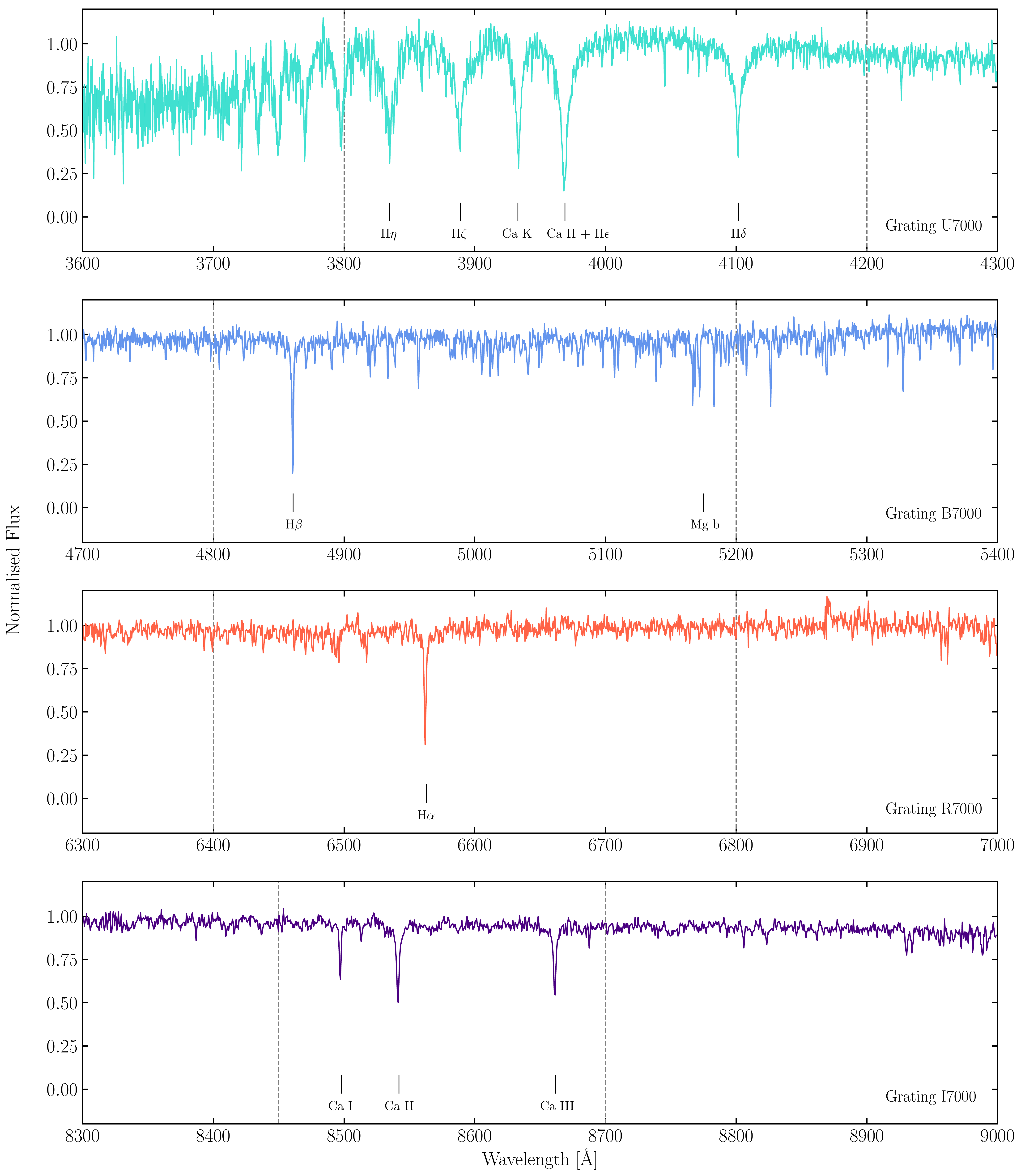}
    \caption{Example spectra (\textit{Gaia} DR2 ID: 6632372787224214912) of an asymptotic giant branch star in NGC 6752 extracted by PampelMuse across four different gratings (U7000, turquoise; B7000, blue; R7000, red; I7000, purple). The grey dashed lines mark the wavelength ranges used to cross-correlate against the PHOENIX template spectra. The wavelength ranges were chosen to avoid telluric lines. Prominent hydrogen and calcium lines are also marked.}
    \label{fig:gratings}
\end{figure*}

As our analysis aims to maximize the number of extracted stars, it will always yield a fraction of spectra with a signal-to-noise ($S/N$) that is not high enough ($\gtrsim 5$\,\AA$^{-1}$) for a meaningful analysis. This problem often affected the spectra of red giant branch stars observed with the U7000 and B7000 gratings, in particular for clusters with high extinctions. 
Example spectra of an asymptotic giant branch star from NGC 6752 (\textit{Gaia} DR2 ID: 6632372787224214912) can be seen in Figure \ref{fig:gratings}. The $S/N$ per pixel is higher than the typical observed spectra, and is as follows: 14.3 (U7000), 24.4 (B7000), 26.1 (R7000), and 29.6 (I7000).

\begin{figure}
    \centering
    \includegraphics[width=0.45\textwidth]{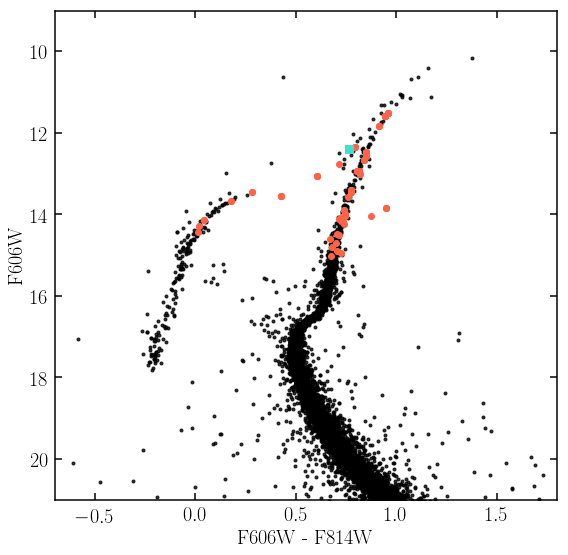}
    \caption{An example colour-magnitude diagram of NGC 6752 with photometry from the UV Legacy Survey \citep{2017AJ....153...19S}. Stars with WAGGS spectra are marked in red. The turquoise square corresponds to the spectrum depicted in Figure \ref{fig:gratings}.}
    \label{fig:cmd}
\end{figure}

For each spectrum, effective temperature ($T$) and surface gravity (log $g$) were determined by over-plotting MIST (MESA Isochrones \& Stellar Tracks) isochrones \citep{2016ApJS..222....8D,2016ApJ...823..102C,2011ApJS..192....3P,2013ApJS..208....4P,2015ApJS..220...15P} on a colour-magnitude diagram (CMD). We assumed an age of 12.59 Gyr for each cluster and adopted the metallicities in the 2010 edition of the \citet{1996AJ....112.1487H,2010arXiv1012.3224H} catalogue. %These assumptions are good enough for the purpose of determining $T$ and log $g$.
An example $m_{\text{F606W}}-m_{\text{F814W}}$ versus $m_{\text{F606W}}$ CMD for NGC 6752 can be seen in Figure \ref{fig:cmd}. The stellar spectra were cross-correlated against a grid of $T$ and log $g$ values corresponding to a synthetic PHOENIX template spectrum \citep{2013A&A...553A...6H}. GC $\upalpha$-element abundances (either [$\upalpha$/Fe] = 0.2 or 0.4) were, in the first instance, taken from \citet{2010A&A...516A..55C}. In the second instance, we used \citet{2016A&A...590A...9D} in the absence of \citet{2010A&A...516A..55C} data. %which is based on \citealt{2005A&A...443..735C} stellar spectra templates. 
For the clusters lacking [$\upalpha$/Fe] measurements, [$\upalpha$/Fe] = 0.4 was used. 

We determined radial velocities using \texttt{FXCOR} in PyRAF \citep{2000ASPC..216...59G,2006HSTc.conf..437G,2001ASPC..238...59D,2012ascl.soft07011S}; a task based on the Fourier cross-correlation method developed by \citet{1979AJ.....84.1511T}. The wavelength ranges used for the cross-correlation were selected in order to include important lines (e.g. Calcium triplet, H$\upalpha$) while avoiding telluric lines. We used wavelength ranges as follows: I7000 grating ($\uplambda\uplambda = 8450-8700$ \AA); R7000 grating ($\uplambda\uplambda = 6400-6800$ \AA); B7000 grating ($\uplambda\uplambda = 4800-5200$ \AA); and U7000 grating ($\uplambda\uplambda = 3800-4200$ \AA), shown by the grey dashed lines in Figure \ref{fig:gratings}.
Since each spectrum was cross-correlated against a grid of PHOENIX spectra, the radial velocity with the highest Tonry and Davis Ratio ($TDR$; a measure of how good the fit is) was used.

\begin{figure*}
    \centering
    \includegraphics[width=\textwidth]{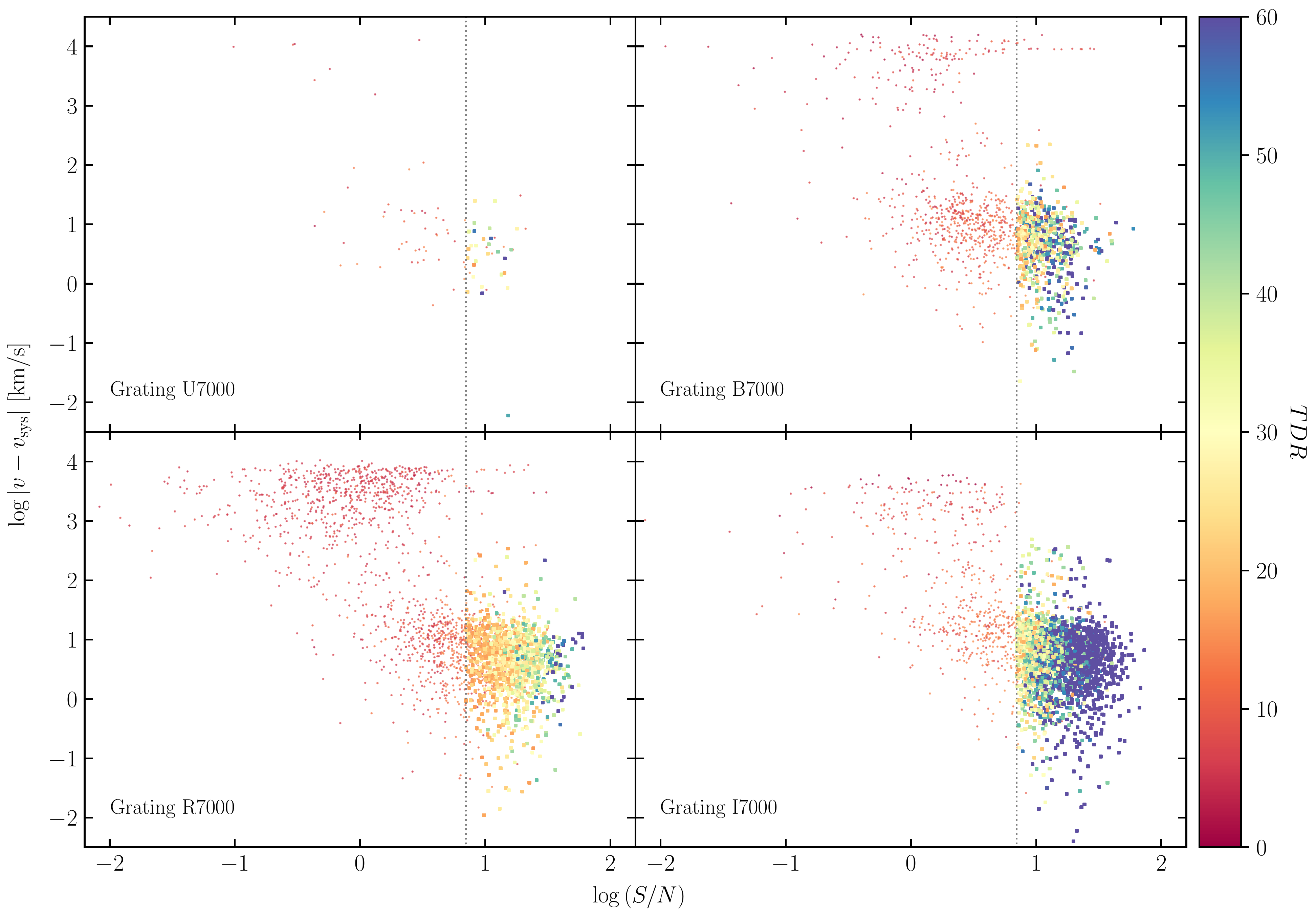}
    \caption{Absolute velocity difference (with respect to the cluster systemic velocity) versus $S/N$ for all spectra for each grating in log-log space. The spectra belong to all 59 clusters and are coloured according to their $TDR$ value. Spectra with $TDR < 16$ (shown as small dots) and $S/N < 7$ (grey dotted lines) are deemed unreliable and have been excluded from further analysis. A maximum of $TDR = 60$ has been applied to better see the spectra at lower values.}
    \label{fig:v_snr}
\end{figure*}

Similar to the method of \citet{2014A&A...566A..58K}, we plot systemic velocities from each grating against $S/N$ in log-log space (Figure \ref{fig:v_snr}), in order to remove unreliable measurements from the RV sample. $S/N > 8$ and $TDR > 13$ yielded 4136 stars for all 59 clusters, whereas $S/N > 7$ and $TDR > 16$ yielded 4258 stars, thus we chose the latter in order to retain the largest number of stars. The grey dotted line in Figure \ref{fig:v_snr} depicts the cut at $S/N = 7$, and spectra shown as small dots have values of $TDR < 16$. The same cut was used for all four gratings. After applying these cuts we are left with a sample of 29, 631, 1169, and 2429 spectra in the U7000, B7000, R7000, and I7000 gratings, respectively. %See Table \ref{table:nstars}.

\begin{figure}
    \centering
    \includegraphics[width=0.45\textwidth]{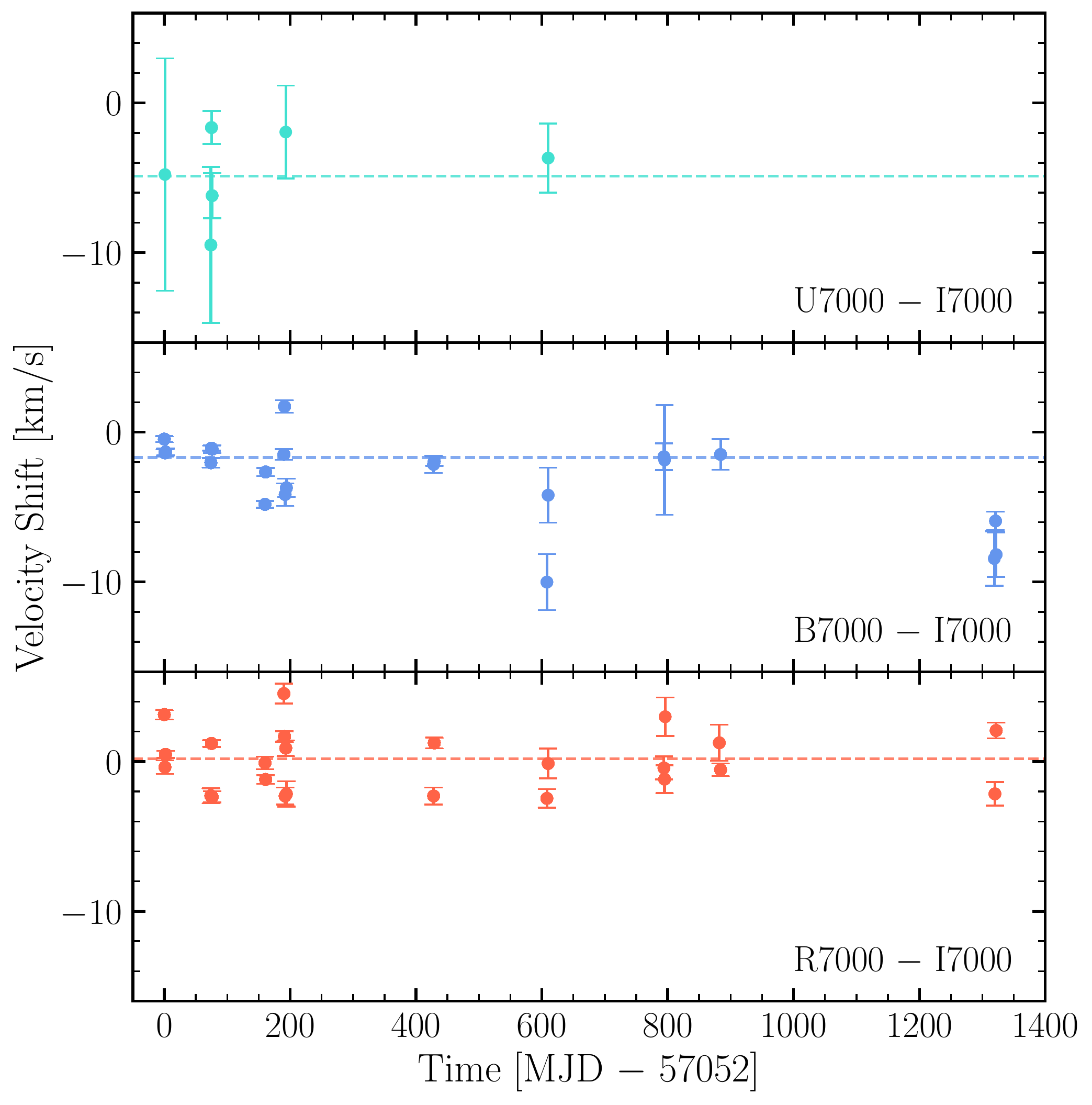}
    \caption{The average shift in velocity in the U7000 (turquoise), B7000 (blue), and R7000 (red) gratings with respect to the I7000 grating. The average velocity shift is shown as a dashed line. Shifts are calculated by taking the differences in velocity measurements that are associated with the same star, and then averaging over all the velocity differences from all stars each night.}
    \label{fig:v_shifts}
\end{figure}

To account for nightly systematic variances between the different gratings, we calculated a shift in velocity between the different gratings, relative to the I7000 grating, per star per night. The I7000 grating was chosen since it contains the greatest number of spectra with the smallest uncertainties. Averaging the shifts for all stars each night, we applied a correction to the radial velocities in the U7000, B7000, and R7000 gratings (Figure \ref{fig:v_shifts}). %We were unable to apply this shift for 0 nights, for which I7000 spectra were missing (which equates to 7.6\%).

After averaging the velocities obtained from the different gratings, we applied sigma clipping ($\upsigma$ = 3) to exclude any outliers in velocity space. For NGC 6496, for which we have very few stars, two clear outliers were still present (with radial velocities differing by $\sim160$ km/s), and were removed by hand. Overall, we use only sufficiently accurate velocity measurements, usually velocity errors less than 1.5--2 km/s, depending on the internal velocity dispersion of the
cluster.

%The velocity dispersion ($\upsigma$) is calculated following the technique of \citet{1993ASPC...50..357P} and influenced by \citet{2013PASP..125..306F} to incorporate an affine-invariant ensemble sampler for Markov chain Monte Carlo (MCMC; \citealt{2010CAMCS...5...65G}).

\subsection{\textit{N}-body models}

We compared the observational data of each globular cluster with a large grid of 2,500 $N$-body simulations. To this end, we ran $N$-body simulations of isolated star clusters --- each containing $N=100,000$ stars --- using the GPU-enabled version of the collisional $N$-body code NBODY6 \citep{2012MNRAS.424..545N}. 
The simulated clusters followed \citet{1962AJ.....67..471K} density profiles with variations of initial concentrations ($0.2 \le c \le 2.5$) and initial radii ($2 \le r_h \le 35$ pc). We also varied the initial mass function of the star clusters from those following an initial \citet{2001MNRAS.322..231K} mass function to clusters highly depleted in low-mass stars. 

All simulations were run up to an age of $T=13.5$ Gyr and final cluster models were calculated by taking 10 snapshots from the simulations centred around the age of each GC. The combined snapshots of the $N$-body clusters were scaled in mass and radius to match the density and velocity dispersion profiles of the observed globular clusters and the best-fitting model was determined by interpolating in the grid of $N$-body simulations. %\textbf{For this work, the models have been extended up to [Fe/H] $= -0.5$ dex for all mass functions, as opposed to the previous value of [Fe/H] $= -1.3$ dex.} 
Further details of the performed $N$-body simulations can be found in \cite{2017MNRAS.464.2174B} and \cite{2018MNRAS.478.1520B}.

\section{Results}
\label{sec:results}

We combine the WAGGS radial velocities with the \citet{2019MNRAS.482.5138B} data and fit the $N$-body models to each cluster. Each set of velocity measurements are shifted to a common velocity by cross-matching the different data sets against each other to correct for zero-point offsets. We then cross-correlate against \textit{Gaia} position measurements to get proper motion information in order to select which stars to use, replacing any original positions with the \textit{Gaia} positions if a match is found. We finally average all velocity measurements for each star and exclude any stars with varying radial velocities.

For the WAGGS data there is the added complication that the stars are located in the crowded cluster centres where \textit{Gaia} measurements are lacking, and all data come with position errors. About 20-100\% of the WAGGS stars have \textit{Gaia} counterparts depending on completeness. We use a small search radius (0.5'') to minimise erroneous cross identifications.

\begin{figure}
    \centering
    \includegraphics[width=0.45\textwidth]{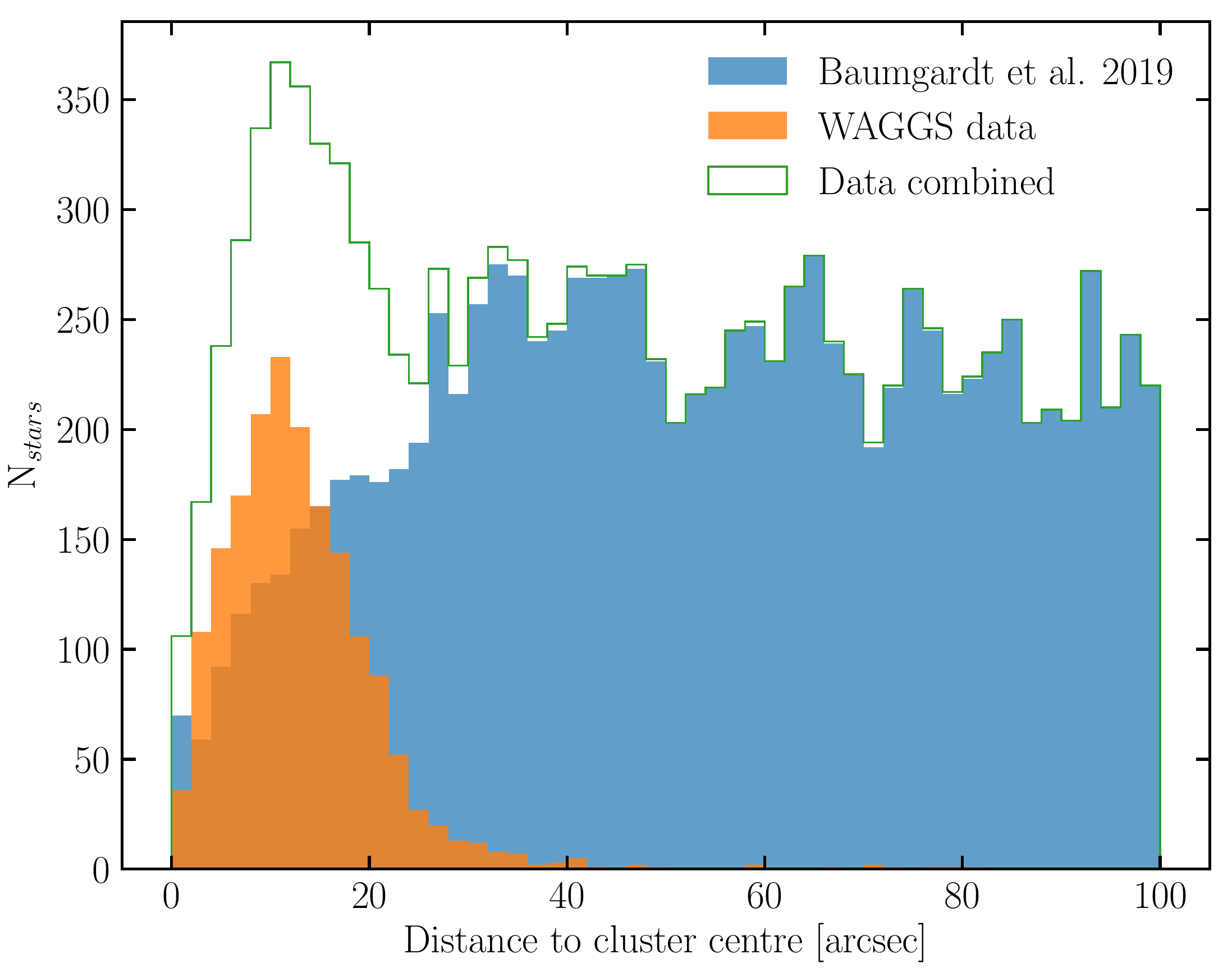}
    \caption{The number of our radial velocity measurements of the 59 Milky Way globular clusters in our sample. We apply a cut at 100 arcsec to emphasize our data at the cluster centre. In blue are the previous values already in the \citet{2019MNRAS.482.5138B} sample. In orange are the measurements that we contribute, the majority of which lie at the cluster centres (mostly within 20 arcsec), significantly increasing the number of velocity measurements in this parameter space. The green outline shows the cumulative sum of the total sample used for $N$-body modelling.}
    \label{fig:hist_r}
\end{figure}

In the end, we increase the \citet{2019MNRAS.482.5138B} dataset with 1,622 WAGGS measurements, within approximately 20 arcsec from the cluster centres (Figure \ref{fig:hist_r}). This is smaller than the half-light radii, where the ratio of observed to half-light radius is $\sim0.3$ on average, and ranges between 0.1 (for the more dispersed) and 0.8 (for the more compact GCs). The dataset can be found online\footnote{\url{https://people.smp.uq.edu.au/HolgerBaumgardt/globular/}}. For 16 of the clusters, the number of velocity measurements has increased by more than 50\%. %In total we have 41,354 velocity measurements, reaching angular distances out to 3,000 arcsec from the cluster centres. 

Velocity dispersion profiles are determined by combining the velocity measurements with \textit{Gaia} proper motions (where available). The stars are ordered according to central distance before calculating velocity dispersion (using a maximum likelihood approach, see \citealt{1993ASPC...50..357P}) and $\upchi^2$ for each member --- those with too large $\upchi^2$ are removed. This process is repeated until a stable solution is found. Velocity errors have been taken into account.

\begin{figure*}
    \centering
    \includegraphics[height=0.334\textwidth]{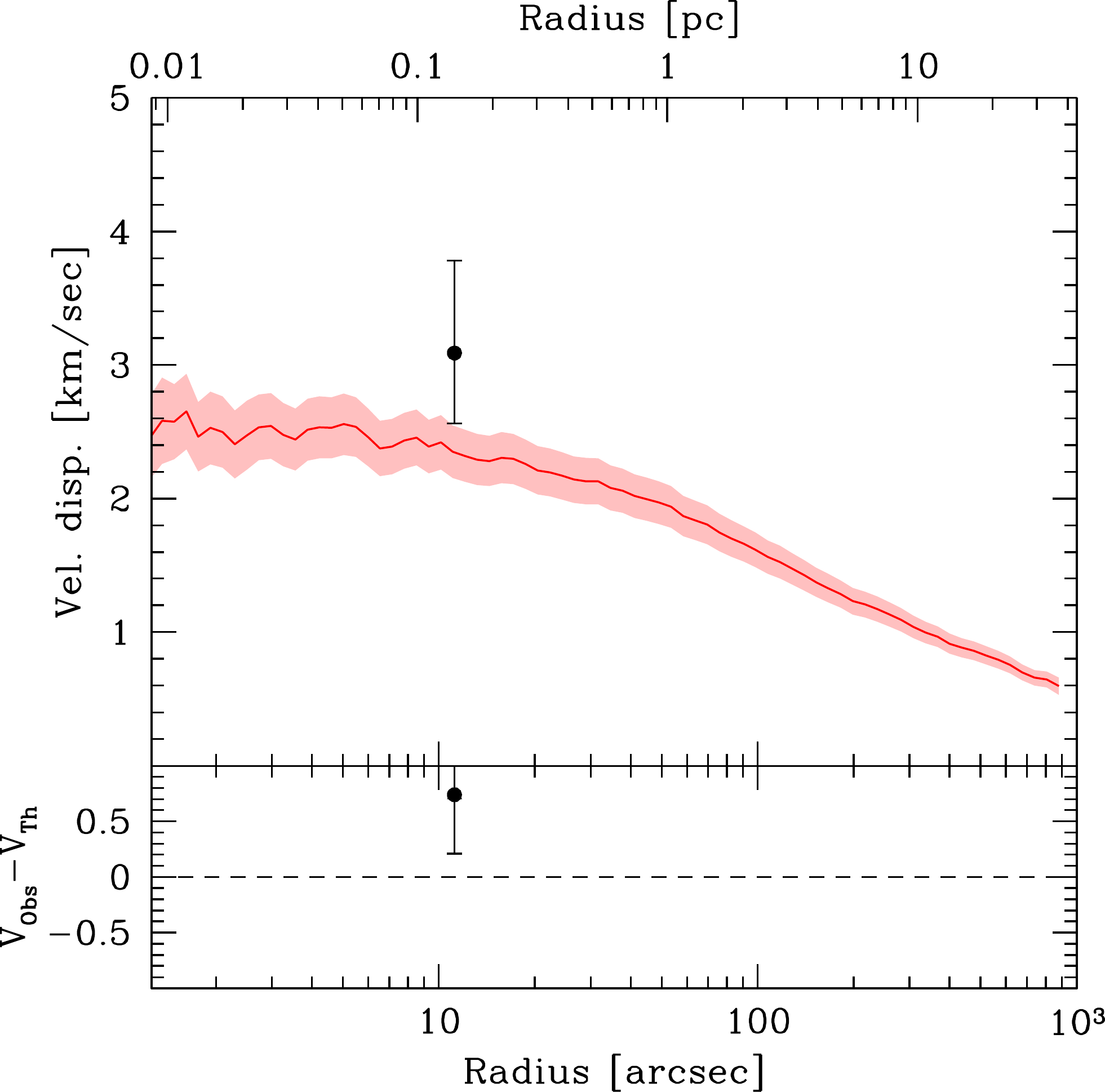}
    \includegraphics[height=0.332\textwidth]{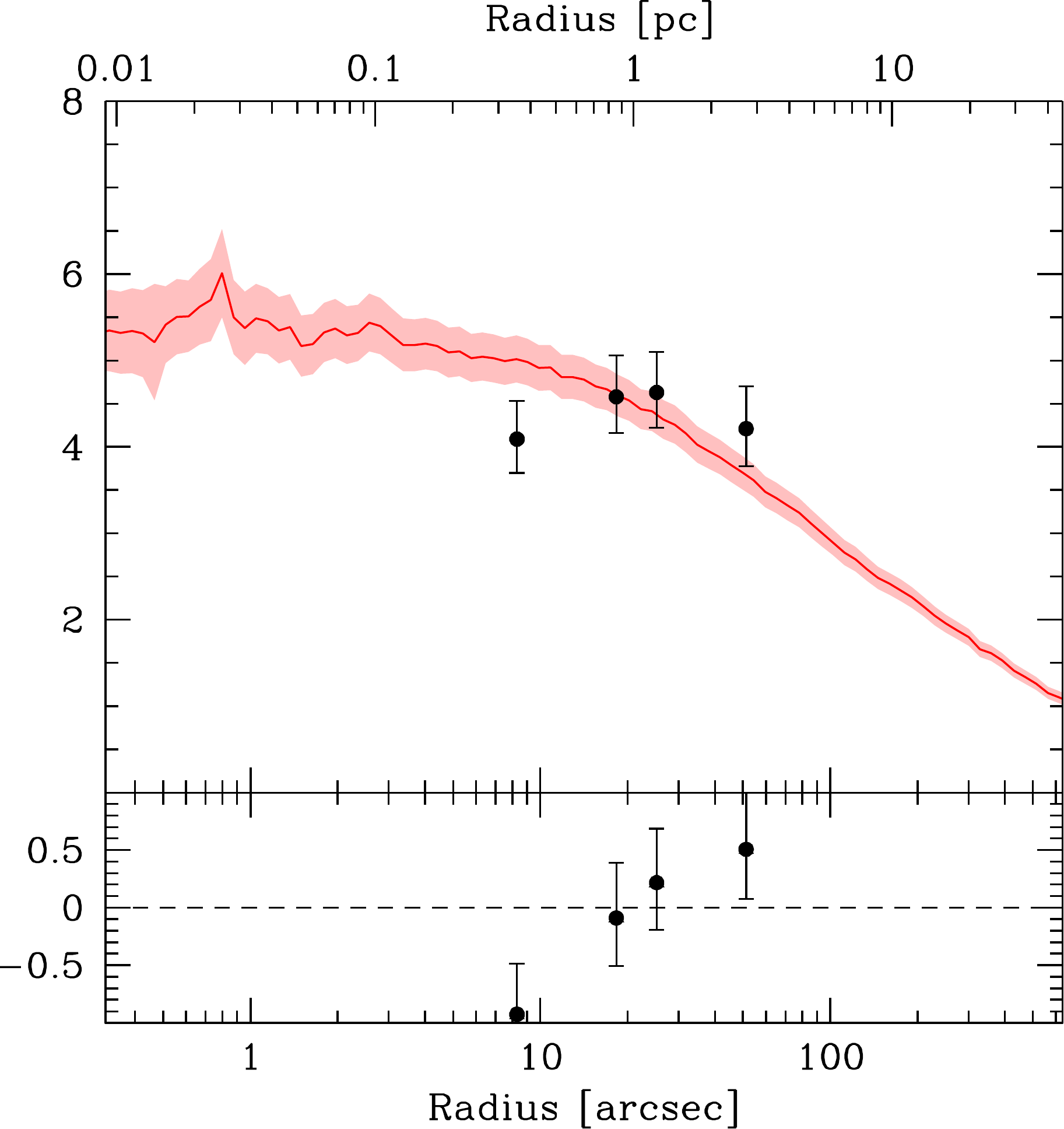}
    \includegraphics[height=0.332\textwidth]{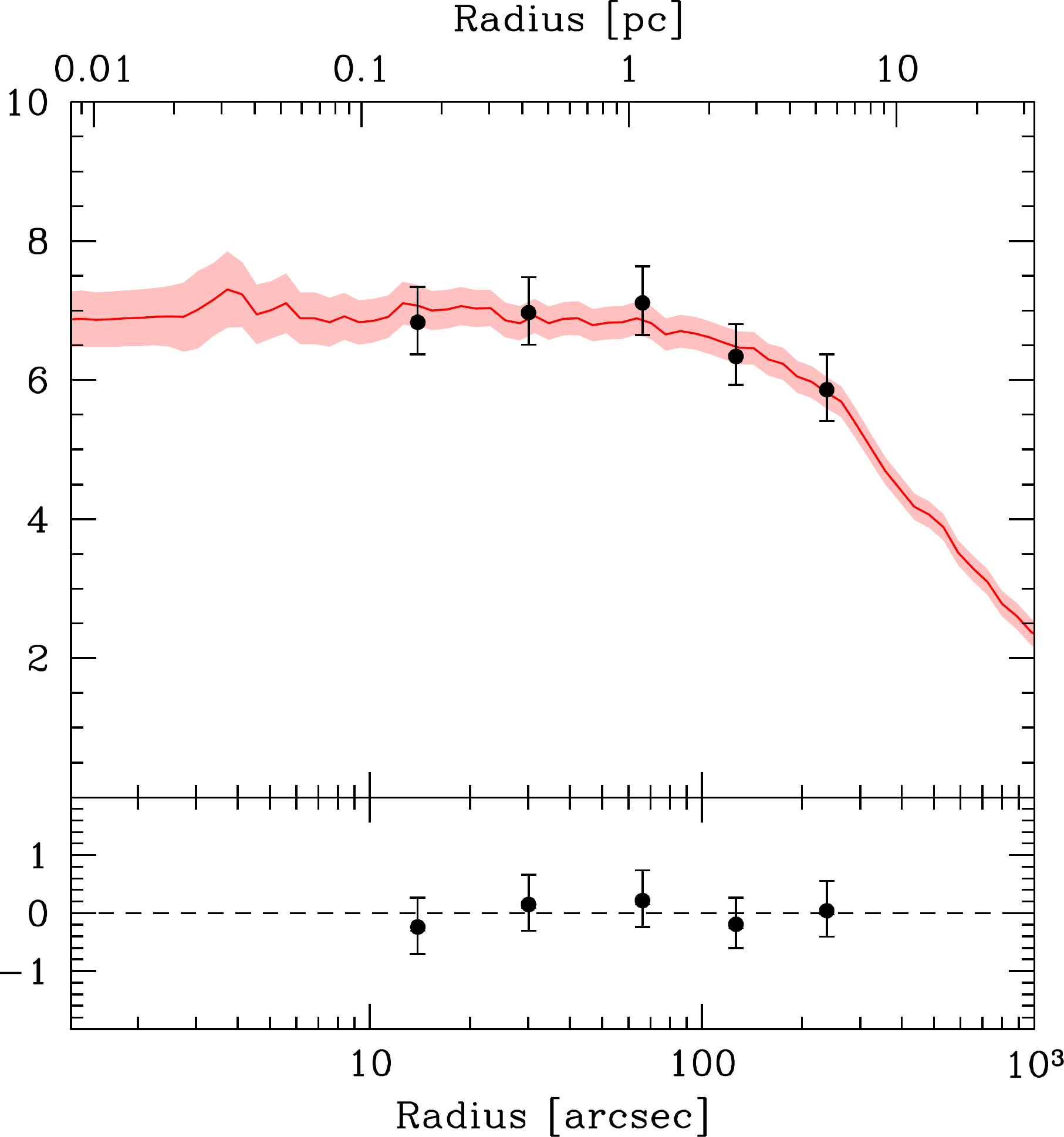}
    \caption{$N$-body fits for NGC 6717 (left), NGC 6528 (middle), and NGC 6553 (right) radial velocities (not including proper motions). The reduced chi-squared values are 3.60, 1.60, and 0.14, respectively. }
    \label{fig:rvfits}
\end{figure*}

We then combine each velocity dispersion profile with the corresponding surface density profile and stellar mass function in order to derive cluster mass. This is done by fitting a grid of $N$-body simulations to the combined data as described in \citet{2017MNRAS.464.2174B,2018MNRAS.478.1520B} --- examples of which can be seen in Figure \ref{fig:rvfits}.

Cluster $M/L_{\mathrm{V}}$ ratios are calculated by taking the mean cluster luminosity from 
\cite{1996AJ....112.1487H,2005ApJS..161..304M,2012AJ....144..126D} in addition to luminosity determinations from \citet{2018MNRAS.478.1520B}. The results are presented in Table \ref{table_gc_params}.

\input{table_gc_parameters.tex}

\subsection{Comparisons to Previous Work}

We compare our derived velocity dispersions and masses with previous work: 20 clusters overlap with the \citet{2015AJ....149...53K} sample; and ten can be compared to \citet{2018ApJ...860...50F}. The central velocity dispersions derived by each study are (King) model predictions based on velocity dispersions further out. For completeness, we also compare the results of \citet{2019MNRAS.482.5138B} to ours. %We do not compare our results with \citet{2015A&A...573A.115L,2018MNRAS.473.5591K} etc. since they have not calculated GC masses (could compare vel. disp?). 
Twelve clusters (NGCs 5946, 6325, 6333, 6342, 6355, 6356, 6380, 6453, 6517, 6642, 6760, 7006) are previously little studied, and thus cannot be compared.

\begin{figure}
    \centering
    \includegraphics[width=0.45\textwidth]{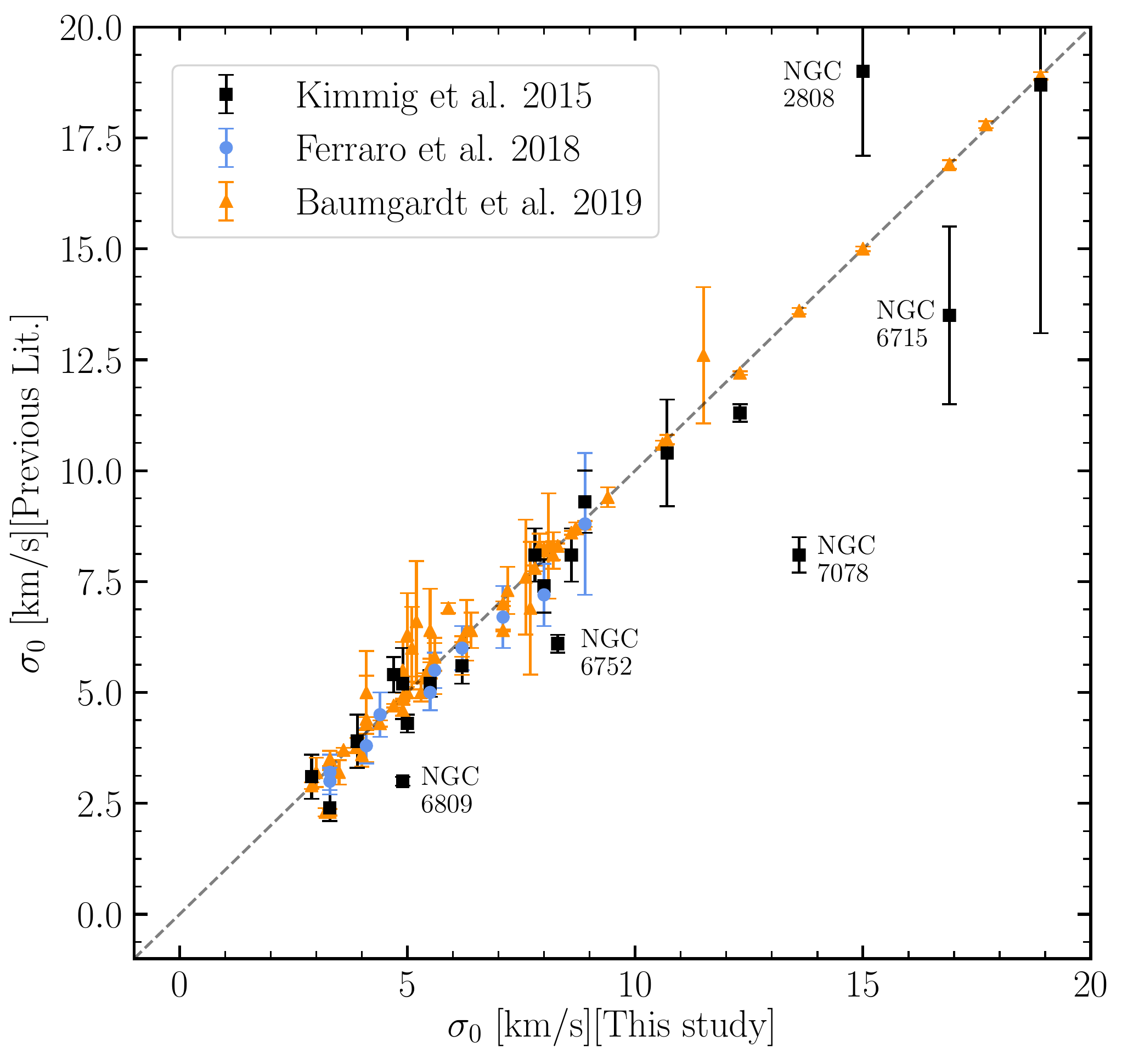}
    \caption{The (1D) central velocity dispersions of 59 WAGGS Milky Way globular clusters compared to \citeauthor{2015AJ....149...53K} (\citeyear{2015AJ....149...53K}, black squares), \citeauthor{2018ApJ...860...50F} (\citeyear{2018ApJ...860...50F}, blue circles), and \citeauthor{2019MNRAS.482.5138B} (\citeyear{2019MNRAS.482.5138B}, orange triangles). Some of the most discrepant GCs from \citet{2015AJ....149...53K} are labelled.}
    \label{fig:sig0}
\end{figure}

Figure \ref{fig:sig0} shows our central velocity dispersions compared to the aforementioned literature. %We also find that the central velocity dispersions of \citet{2015AJ....149...53K,2018ApJ...860...50F} tend to be lower than ours. This may be due to contamination by the background light of the GC; this would cause central velocity measurements to shift towards the mean cluster velocity, which would have a stronger effect at the cluster centre. Whereas IFU spectroscopy is not susceptible to this effect \citep{2013A&A...549A..71K}.
The most significant discrepancies lie with the results of \citet{2015AJ....149...53K}, particularly for NGCs 2808, 6715, and 7078. We further observe a trend for the central velocity dispersions of \citet{2018ApJ...860...50F} to be lower than ours. As their results are based on multi-object spectroscopy, this could be caused by differences in the average distances of the sample stars to the cluster centres. Also, contamination from nearby sources or the unresolved cluster light can be an issue near the centres. However, as our analysis explicitly accounts for these effects, we are confident that our values are robust against such contamination.

\begin{figure}
    \centering
    \includegraphics[width=0.45\textwidth]{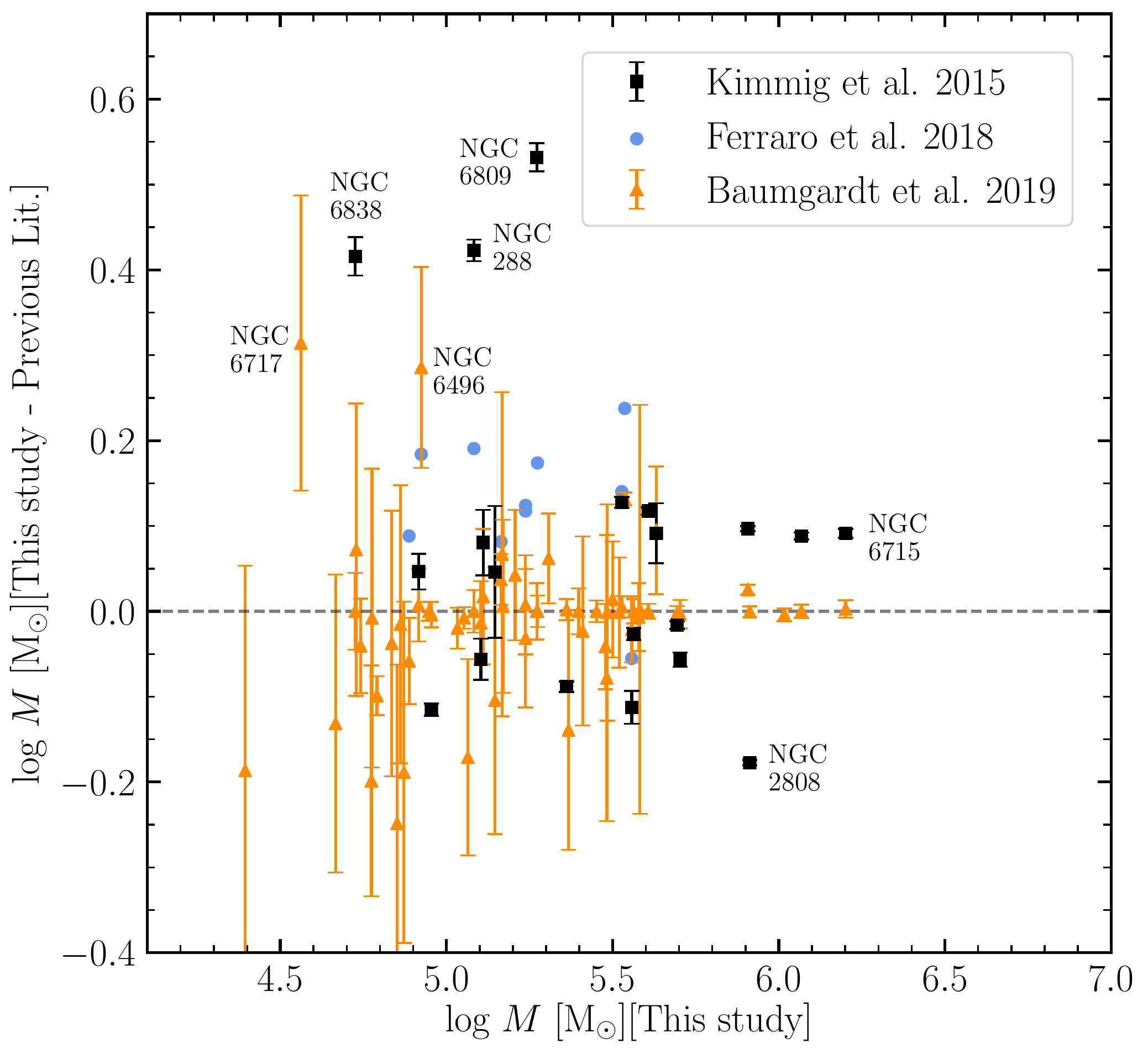}
    \caption{The masses of Milky Way globular clusters derived here compared to \citet{2015AJ....149...53K,2018ApJ...860...50F,2019MNRAS.482.5138B}, the same as in Figure \ref{fig:sig0}. Some of the most discrepant \citet{2015AJ....149...53K,2019MNRAS.482.5138B} GCs are labelled.}
    \label{fig:mass}
\end{figure}

Figure \ref{fig:mass} shows our derived cluster masses compared to \citet{2015AJ....149...53K,2018ApJ...860...50F,2019MNRAS.482.5138B}. As before, three of the \citet{2015AJ....149...53K} cluster values are the most discrepant, although for different GCs: NGCs 288, 6809, and 6838. The majority of the \citet{2018ApJ...860...50F} masses are shifted towards lower masses, which appears to be a consequence of the on-average lower dispersion measurements in the study of \citet{2018ApJ...860...50F}.

%NGCs 5946, 6325, 6333, 6342, 6355, 6356, 6380, 6453, 6517, 6642, 6760, 7006.

\begin{figure}
    \centering
    %\includegraphics[width=0.45\textwidth]{kimmig.png}
    %\caption{The M/L of Milky Way globular clusters versus metallicity. 20 GCs from \citet{2015AJ....149...53K} (red) are compared to this study (teal).}
    \label{fig:kimmig}
\end{figure}

\begin{figure}
    \centering
    %\includegraphics[width=0.45\textwidth]{kimmig_logm.png}
    %\caption{The mass-to-light ratio of Milky Way globular clusters versus log mass. 20 GCs from \citet{2015AJ....149...53K} (red) are compared to this study (teal).}
    \label{fig:kimmig_logm}
\end{figure}

\begin{figure}
    \centering
    \includegraphics[width=0.47\textwidth]{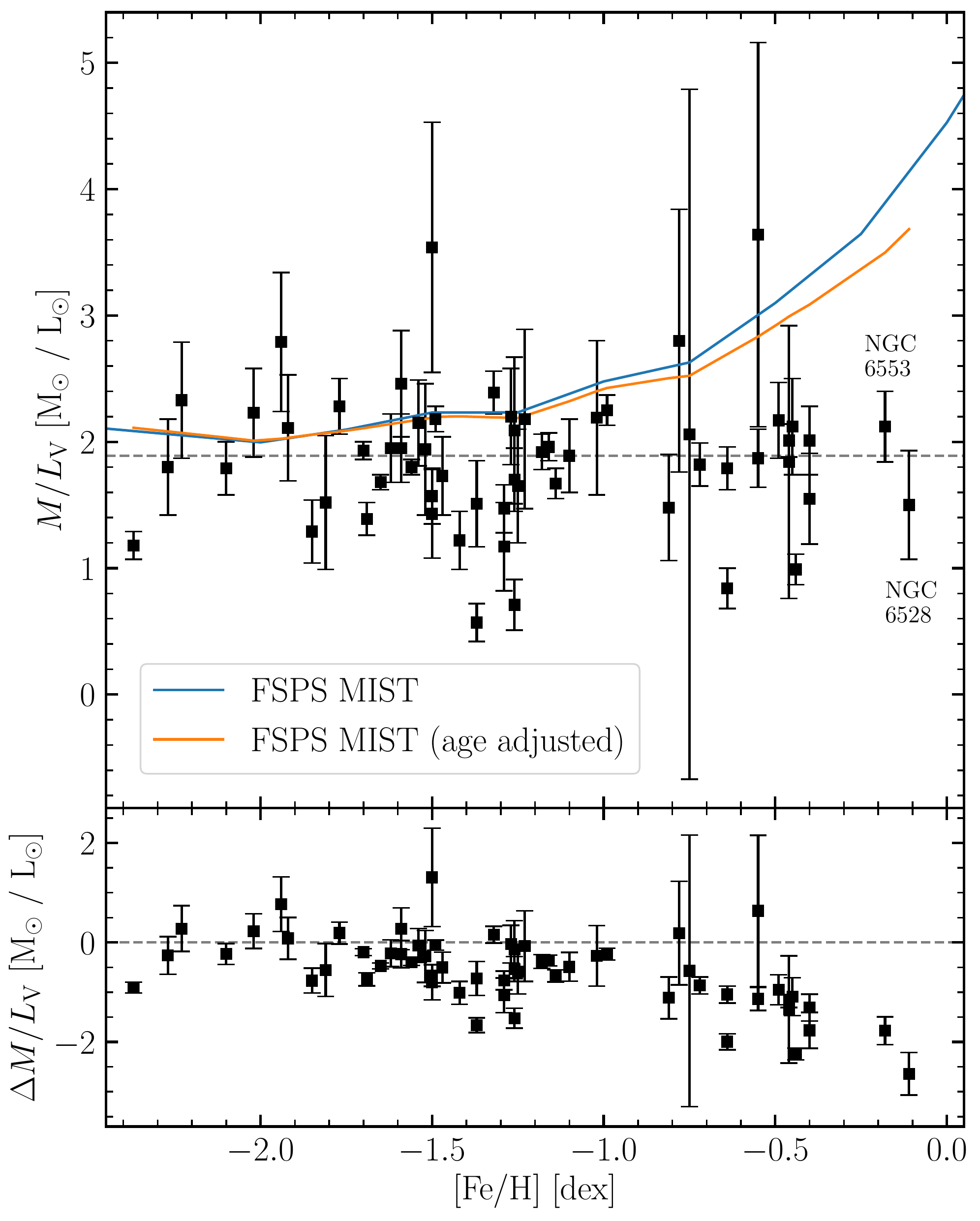}
    \caption{The mass-to-light ratio of Milky Way globular clusters versus metallicity. Over-plotted are the original FSPS MIST model at 12.59 Gyr (blue) and a 2D interpolated version which accounts for the age-metallicity relation derived by \citet{2019MNRAS.486.3180K} (orange). The median $M/L$ ratio is shown as a grey dashed line. The difference in $M/L$ between each GC and the original FSPS MIST model is shown underneath.}
    \label{fig:feh_ml}
\end{figure}

\section{The mass-to-light ratio}

The $M/L$ depends on the proportion of high- to low-mass stars within a cluster. Massive stars contribute most of the light within a cluster, whereas the low-mass stars and stellar remnants (white dwarfs, neutron stars and black holes) contribute most of the mass.
As such, the $M/L$ provides an important insight into stellar evolution and the stellar mass function, and is a very useful tool for checking and constraining SSP models. 

We plot the $M/L_\mathrm{V}$--[Fe/H] relation of 59 Milky Way globular clusters, reaching higher metallicities --- [Fe/H] $> -0.4$ dex --- than seen before (Figure \ref{fig:feh_ml}). The median $M/L_\mathrm{V}$ is shown as a grey dashed line (1.9 M$_{\sun}$/L$_{\sun}$). We find no trend with [Fe/H], with Spearman's rank correlation coefficient, $\rho=-0.01$. However, even after accounting for observational errors, a cluster-to-cluster scatter is clearly visible. %(standard deviation = 0.44).
In particular, we note five outliers:
\begin{enumerate}
\item NGCs 6342 and 6453 ($M/L_\mathrm{V} \sim 3.6$ $\text{M}_{\odot}/\text{L}_{\odot}$) both have large uncertainties, which can be explained by uncertain total cluster luminosities. Newly derived luminosities (Baumgardt, in prep) decrease the $M/L_\text{V}$ of these clusters.
\item NGCs 6355, 6637, and 6642 have low dynamical $M/L_\mathrm{V}$ ($< 0.9$ $\text{M}_{\odot}/\text{L}_{\odot}$) and small uncertainties, also likely due to unreliable luminosities and underestimated errors. Baumgardt (in prep) finds more reliable luminosities, which lead to $M/L_\text{V} > 1$ for all four clusters. Determining accurate luminosities is difficult due to close proximity to the bulge of all three clusters (in galactocentric radius, $1.0 < R_{\text{GC}} < 1.7$ kpc).
\end{enumerate}

In addition to the above, measuring $M/L$ itself is known to be model-dependent, often with a scatter of 20\% or more (e.g. \citealt{2017MNRAS.468.4429Z}, Table 2, and \citealt{2017ApJ...844..167B}, Table 8). However, such errors should not correlate with metallicity; while our errors do not take into account systematic errors introduced by the $N$-body models, it is unlikely that they will counterbalance systemic offsets at high metallicity.

We compare our observations with theoretical predictions by over-plotting stellar population models (Figure \ref{fig:feh_ml}; blue line). The models are calculated using version 3.0 of the Flexible Stellar Population Synthesis (\textsc{FSPS}) code \citep{2009ApJ...699..486C, 2010ApJ...712..833C}, MIST isochrones \citep{2016ApJS..222....8D,2016ApJ...823..102C,2011ApJS..192....3P,2013ApJS..208....4P,2015ApJS..220...15P}, a \citet{2001MNRAS.322..231K} IMF and the MILES spectral library \citep{2006MNRAS.371..703S}.
We also tested the UV-extended E-MILES stellar population models from \citet{2016MNRAS.463.3409V}, using both the BaSTI \citep{2004ApJ...612..168P} and Padova \citep{2000A&AS..141..371G} theoretical isochrones. We find good agreement between all predicted $M/L_\mathrm{V}$ ratios despite using different isochrones and stellar libraries; they all agree within $<10$ \%.

SSP models with fixed IMFs predict that $M/L$ increases with metallicity in the V-band, as a result of line blanketing (e.g. \citealt{2003MNRAS.344.1000B,2005MNRAS.362..799M}). However, as we have extended the range of $M/L_\mathrm{V}$ versus [Fe/H] (Figure \ref{fig:feh_ml}) to include two clusters --- NGCs 6528 and 6553 --- at higher metallicity, it is clear to see that the $M/L_\mathrm{V}$ ratio remains constant.
%\footnote{We note that we have not included NGC 6553 photometry from \cite{2001AJ....121.2618B}, which would suggest that the mass function is less evolved than assumed by our $N$-body model. This would increase the $M/L_\mathrm{V}$ and bring it in better agreement with the SSP prediction, highlighting the potential impact of the assumed mass function on the $M/L_\mathrm{V}$.} 
The models and observations are well matched at lower metallicities, until [Fe/H] $\sim -1.0$ dex when they begin to diverge.

\subsection{Comparisons to previous work}

Since \citet{2009AJ....138..547S,2011AJ....142....8S} first showed the discrepancy between the $M/L$ ratios of stellar population models and globular clusters within M31, several other studies have focused their attention on Galactic GCs  \citep{2015AJ....149...53K,2015ApJ...812..149W,2017MNRAS.464.2174B,2019ApJ...871..159V}.

\citet{2015AJ....149...53K} studied the $M/L$ of 25 Galactic GCs, with metallicities going up to $-0.46$ dex. They compared their observations to SSP models at an age of 13 Gyr \citep{2013A&A...558A..14M}, and found no evidence for an increase in $M/L$ with metallicity.
\citet{2015ApJ...812..149W} used $HST$ proper motions of \cite{2014ApJ...797..115B} to determine dynamical $M/L$ ratios of 15 MW clusters up to $\sim -0.5$ dex. They observed the same decrease in $M/L$ as \citet{2011AJ....142....8S} towards the metal-rich end. When compared to the $M/L$ population-synthesis estimates of \citet{2005ApJS..161..304M}, the metal-poor clusters agree, but offsets exist for the clusters above [Fe/H] = $-1.0$ dex.
The work of \citet{2017MNRAS.464.2174B} %compare their results against four different SPS models: they predict $M/L$ ratios for a Kroupa IMF \citep{2001MNRAS.322..231K} according to PARSEC, BaSTI, and Dartmouth isochrones; they also predict the $M/L$ ratio for a Chabrier IMF \citep{2003PASP..115..763C} according to PARSEC isochrones. Given that they 
included five GCs with  $-1.0 <$ [Fe/H] $< -0.3$ dex, %of which their $\Upsilon_{\text{Obs}}$/$\Upsilon_{\text{Kroupa}}$ ratios lie within the range of values seen at lower metallicities, they 
but due to large uncertainties they are uncertain if the SSP models are in fact very different to the observations at higher metallicities. 
Finally, \citet{2019ApJ...871..159V} studied the $M/L$ of Ultra Compact Dwarfs (UCDs). The authors showed that four UCDs --- which have roughly solar metallicity --- also lie at a $M/L$ below the theoretical prediction, after accounting for the impact of their supermassive black holes. 

%As we have now extended this work to include clusters at higher metallicities, it is now very clear that there lies a discrepancy, which we calculate to be on the order of x. This is a different result to \citet{2011AJ....142....8S,2015AJ....149...53K}, who find that the $M/L$ ratio decreases with [Fe/H].
%This is puzzling, since theoretical predictions made by stellar population models predict that the $M/L$ ratio should increase with [Fe/H]. 
%Several efforts have been made to try to understand where the discrepancy originates.

We find that our results remain consistent with previous work and add further evidence to the discrepancy between stellar population models and observations. The SSP models that we use are typical of those used in previous literature.
However, it is well known that other factors like age or the fraction of compact remnants will have an effect on the $M/L$ of a cluster. We now explore these effects in more detail, in reference to Galactic GCs. %and neglect effects due to dynamical evolution, such as two-body relaxation, mass segregation, and evaporation. In the following sections, 

\begin{figure}
    \centering
    \includegraphics[width=0.47\textwidth]{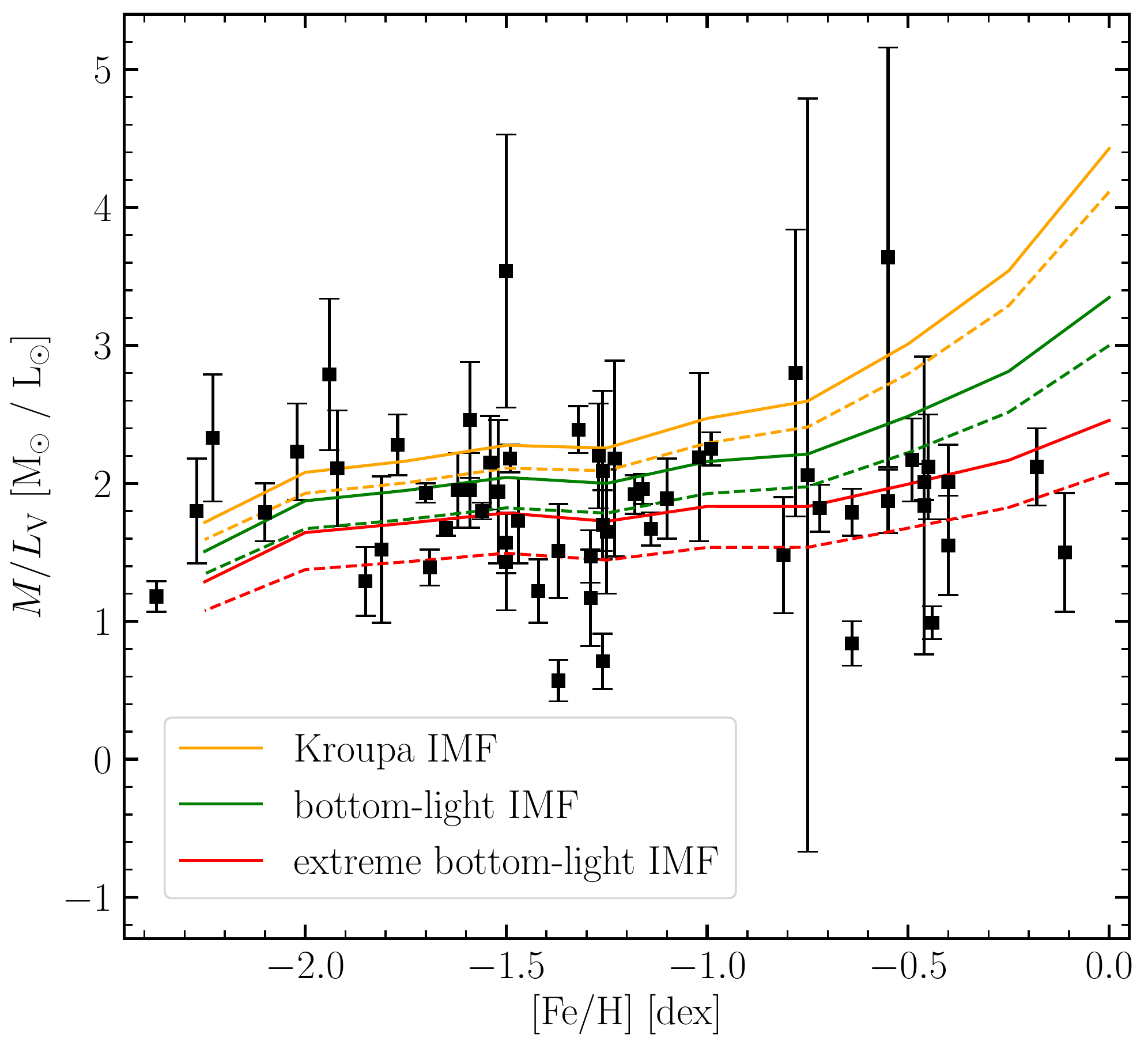}
    \caption{$M/L$ of Milky Way globular clusters versus [Fe/H] for a Kroupa IMF (yellow), bottom-light mass function (green), and an extreme bottom-light mass function (red). The solid lines represent 100\% retention of remnants (i.e. NSs and BHs), compared to 0\% remnant retention (dashed lines).}
    \label{fig:feh_ml_imf_rem}
\end{figure}

\subsection{Cluster ages}

%Flexible SPS (FSPS) model of \citet{2006MNRAS.371..703S,2009ApJ...699..486C,2010ApJ...712..833C} in combination with , and a \citet{2001MNRAS.322..231K} IMF.

Older clusters have a larger fraction of evolved stars compared to younger clusters, increasing the light emitted by a GC, and thus lowering $M/L$. Since the most metal-rich GCs in the MW are $\sim 2.5$ Gyr younger than the most metal-poor GCs (e.g. \citealt{2013ApJ...775..134V}), we expect that $M/L$ will decrease at higher metallicity. 
\citet{2017ApJ...839...60H} examined an empirical relation between age and metallicity and the effect on $M/L$ for M31 GCs but required the most metal-rich GCs in M31 to have significantly younger ages than is observed \citep[e.g.][]{2011AJ....141...61C}. %However, as \citet{2019ApJ...871..159V} states, they use very young ages (down to 1 Gyr) which do not concur with the observed ages of metal-rich M31 clusters (> 5 Gyr). 

We follow a different approach, taking the age-metallicity relation derived by \citet{2019MNRAS.486.3180K} to predict the age of each GC based on its metallicity, and then interpolate to determine the empirical $M/L$ ratio (Figure \ref{fig:feh_ml}; orange line). This decreases the $M/L$ ratio as expected, particularly from [Fe/H] = $-1.0$ dex and above, but not enough to account for the difference between the models and observations at the metal-rich end i.e. the influence of age is minimal. We find that the GCs would need to be as young as $\sim 6$ Gyr in order for the models to agree, similar to what \citet{2019ApJ...871..159V} find for UCDs, but this is several Gyr younger than the ages measured for metal-rich Galactic GCs (e.g. \citealt{2013ApJ...775..134V}). As more observations of younger LMC and SMC clusters become available, it will be possible to see how their $M/L$ relates to older MW GCs, which may shed more light on the effect of cluster age.

%\citet{2017ApJ...839...60H} too tested models for younger ages ($5-6$ Gyr) which did produce a M/L which agrees with observed M/L, but this age is younger than what 

\subsection{Mass function variations}

The $M/L$ ratio of a stellar population strongly depends on its present day mass function (PDMF). 
Dynamical evolution or a varying IMF could alter the proportions of high- or low-mass stars present within a cluster, therefore changing the PDMF in comparison to a Kroupa IMF. 
%For example, fewer low-mass stars (bottom-light) or more high-mass (e.g. RGB/AGB) stars (top heavy) will both work to decrease $M/L$, whereas more remnants and dwarfs (bottom-heavy) or fewer high-mass stars (top-light) will increase $M/L$. 
%A bottom light mass function will have a lower $M/L$ as will a population with a top light mass function (a mass function with fewer stellar remnants compared to a canonical IMF). A population can have a bottom light mass function but still have a $M/L$ heavier than that of a population with canonical IMF if it has sufficiently top heavy mass function.

\citet{2009AJ....138..547S,2011AJ....142....8S} argued that the $M/L$-[Fe/H] relation cannot be due to dynamical effects since the metal-rich and metal-poor GCs of M31 are of similar mass and size. Instead they proposed different IMFs for the metal-rich and metal-poor clusters to try and explain the discrepancy.
\citet{2016ApJ...826...89Z} looked at the effects of a varying IMF for M31 GCs in more detail. They found that a top-heavy IMF, in combination with other effects e.g. dynamical evolution and remnant retention (see Section 5.4 for more detail) improved the agreement between SSP models and $M/L$ observations. 

Here we explore the effect of introducing a bottom-light mass function on the SSP models. Using the MIST isochrones, C3K spectral library (Conroy et al. in prep), and their synthetic $V$-band photometry, we calculate models with three different PDMF (Figure \ref{fig:feh_ml_imf_rem}; solid lines):
\begin{enumerate}
 \item Kroupa IMF, $\alpha = -1.3$ (M < 0.5 M$_{\odot}$), $\alpha = -2.3$ (M > 0.5 M$_{\odot}$) [yellow]; 
 \item bottom-light mass function, $\alpha = -1.3$ (M < 1.0 M$_{\odot}$), $\alpha = -2.3$ (M > 1.0 M$_{\odot}$) [green]; 
 \item extreme bottom-light mass function, $\alpha = -0.3$, (M < 1.0 M$_{\odot}$), $\alpha = -2.3$ (M > 1.0 M$_{\odot}$) [red]. 
\end{enumerate}
For all models we integrate the mass function between 0.08 and 100 M$_{\odot}$ and use the same prescription for remnant masses as used by \citet{1993ApJ...416L..49R}. 

Only the $M/L$ of the third model are low enough to replicate the observations above $-1.0$ dex, however, there is strong evidence that it is nonphysical for the IMF to vary so strongly with [Fe/H] \citep{2010ARA&A..48..339B}. A more detailed study of the influence of the mass function is left for future work. Next, we look to the retention of remnants as another possible effect on the $M/L$. 

\subsection{Remnant retention}

As stars evolve into remnants, $M/L$ increases --- stellar remnants continue contributing to the mass but no longer to the luminosity of a globular cluster. The strength of this effect depends on the number of remnants that are retained throughout a cluster's lifetime. It is expected that at least some proportion of compact remnants (black holes and neutron stars) will either receive a velocity kick when they form and become immediately ejected, or will eventually be expelled from the cluster as a result of dynamical evolution (e.g. \citealt{2018ApJ...864...13W,2019ApJ...881...75K}, and references therein). These compact remnants are removed in conjunction with the evaporation of low-mass stars, a result of mass segregation. 

Previous work has shown that a spread in the retention fraction of compact remnants --- along with metallicity-dependent mass segregation effects --- can explain the $M/L$-[Fe/H] discrepancy of MW clusters \citep{2015MNRAS.448L..94S}. Other work has found that by reducing the remnant retention fraction to 30\% (along with a top-heavy IMF), the discrepancy between the models and observations decreased \citep{2016ApJ...826...89Z}. The latter study explored this further by introducing a dependence on cluster mass and metallicity, which improved their SSP predictions for M31 GCs.

We explore the effect of removing remnants upon the SSP models, by adapting the models from Section 5.3 which follow 100\% black hole (BH) and neutron star (NS) retention. We calculate the equivalent models for the most extreme case: 0\% BH+NS retention in the range $-2.25 <$ [Fe/H] $< 0.00$ for each mass function, at an age of 12.59 Gyr (Figure \ref{fig:feh_ml_imf_rem}; dashed lines). As expected, this has the effect of decreasing $M/L_\mathrm{V}$ for all metallicities. At [Fe/H] $= 0$ dex, the effect of removing all BHs and NSs is $\sim 7\%$ for a Kroupa IMF [case (i)] or $\sim 20\%$ for the bottom-light mass functions [cases (ii) and (iii)]. 

While a 0\% BH+NS retention fraction alone is not enough to explain the discrepancy, we clearly observe some effect. In order to better approximate the percentage of remnants retained within a cluster, it is crucial for future studies to determine the remnant fraction, particularly at the metal-rich end.

\subsection{Other possible effects}

As we have shown in the previous sections, accounting for the younger ages of an increasingly metal-rich cluster population and testing for the effect of removing remnants both serve to decrease $M/L$. Yet each effect alone is not enough to explain the discrepancy between the observed and theoretically predicted $M/L$. Therefore, a combination of the explored effects, or some other effects must be at play.

It is already known that a strong correlation is observed between the dynamical age of a globular cluster and the low-mass slope of its mass function (e.g. \citealt{2017MNRAS.471.3668S}). In other words, mass segregation leads to a preferential loss of low-mass stars in the Galactic tidal fields, impacting on the evolution of a cluster. Therefore, since metal-rich MW clusters are only found in the bulge where they are subject to significant dynamical effects, we expect that they will lose a larger fraction of low-mass stars and have a lower $M/L$. Our $N$-body models take tidal fields into account by adjusting the initial mass function; we note that this approach may be limited in its accuracy.

\citet{2015MNRAS.448L..94S} found that mass segregation leads to a bias in the determination of $M/L$, which was not taken into account by the models of \citet{2011AJ....142....8S}. This bias is accounted for by our $N$-body models, however, so it is unlikely that this could explain the discrepancy between the observed $M/L$ ratio and SSP models. This effect may also explain why \citet{2011AJ....142....8S} find a decrease in $M/L$ with metallicity, while we see a flat relation.

The dynamical evolution of a cluster also depends on its radius; GCs with small radii evolve more quickly. This is true both for Milky Way clusters and their extragalactic counterparts (see \citealt{2013MNRAS.436.1172U} and references therein).
Since metal-rich GCs have smaller radii on average --- smaller galactocentric radii leads to stronger tidal fields --- metal-rich GCs should lose more low-mass stars, i.e. they will have a more bottom-light mass function compared to metal-poor GCs with the same mass. Thus we would expect metal-rich GCs to have lower $M/L$ compared to the predictions of a stellar population calculated using the IMF alone. 

The internal configuration of a cluster itself is also of primary importance (e.g. density, initial tidal filling configuration, relaxation time). More compact clusters evolve faster dynamically, decreasing their $M/L$ accordingly. \citet{2017MNRAS.469.4359B} suggest that possible differences in internal configurations between metal-rich and metal-poor clusters (e.g. metal-rich clusters being more dense) could produce a different dynamical evolution of the $M/L$, therefore decreasing the $M/L$ of metal-rich clusters. This effect could add up to the evolution due to the tidal field.

Another possible cause could lie with the shortcomings of the $N$-body models e.g. binaries or rotation. However, we expect binaries and rotation to be more effectively destroyed in metal-rich clusters, since the inner metal-rich GCs are more compact. In future work, it would interesting to make a more detailed comparison between clusters in the bulge and in the halo, as a more robust test for dynamical effects upon GCs.

Finally, we rule out dark matter as a possible effect, as there is no evidence for dark matter in any MW globular cluster thus far \citep{2009MNRAS.396.2183S,2011ApJ...738..186I}.

\begin{figure}
    \centering
    %\includegraphics[width=0.45\textwidth]{m_ml.png}
    %\caption{The mass-to-light ratio of Milky Way globular clusters versus log mass. The grey dots depict all 156 Milky Way GCs in \citet{2019MNRAS.482.5138B}. The 59 GC results from this study are shown as black squares.}
    \label{fig:m_l_logm}
\end{figure}

\section{Summary}

In this work, we have determined 1,622 radial velocity measurements across 59 Milky Way clusters. The majority of the stars are located in the centres of GCs (within 20 arcsec), extending prior work to a new region of parameter space. Incorporating this new data, we recalculate dynamical parameters (i.e. central velocity dispersions and masses). Importantly, two new MW clusters with [Fe/H] $\gtrsim -0.4$ dex have been introduced --- where previous data had been lacking --- allowing us to better understand the $M/L_\mathrm{V}$-[Fe/H] relation at the metal-rich end.

Our results confirm previous work by \citet{2017MNRAS.464.2174B}; the mass-to-light ratio of GCs does not change with metallicity for Milky Way clusters. Subsequently, we now see an even greater divergence with SSP models, strengthening the concern that we need to decipher where this discrepancy originates. 
Having looked at possible explanations, it seems reasonable to assume that dynamical effects have a significant part to play in the solution. Moreover, globular clusters undergo internal dynamical evolution  much more rapidly than galaxies, and is further accelerated for the metal-rich clusters located in the bulge. This suggests that it is not straightforward to compare globular clusters with galaxies, particularly towards higher metallicities. Either more equivalent calibrators need to be found, or we need to improve our understanding of the dynamical processes so that they can be accounted for in the models. Further work in this area has the potential to reveal a new understanding of the dynamics of globular clusters.

\section*{Acknowledgements}

We thank the anonymous referee for their helpful and insightful comments. Nate Bastian, Christopher Usher, and Sebastian Kamann gratefully acknowledge support from the European Research Council (ERC-CoG-646928, Multi-Pop).
Nate Bastian also gratefully acknowledges financial support from the Royal Society (University Research Fellowship), and Pierluigi Cerulo acknowledges the support of a ALMA-CONICYT grant no. 31180051.

Additionally, we acknowledge the use of archival observations made with the NASA/ESA Hubble Space Telescope, obtained at the Space Telescope Science Institute, which is operated by AURA, Inc., under NASA contract NAS 5-26555.
We also used data from the European Space Agency (ESA) mission {\it Gaia} (\url{https://www.cosmos.esa.int/gaia}), processed by the {\it Gaia} Data Processing and Analysis Consortium (DPAC, \url{https://www.cosmos.esa.int/web/gaia/dpac/consortium}). Funding for the DPAC has been provided by national institutions, in particular the institutions participating in the {\it Gaia} Multilateral Agreement.
 
Finally, we note that this work made use of \textsc{numpy} \citep{numpy}, \textsc{scipy} \citep{scipy}, \textsc{matplotlib} \citep{matplotlib}, \textsc{astropy} \citep{2013A&A...558A..33A}, and \textsc{pyraf}, a product of the Space Telescope Science Institute.

%%%%%%%%%%%%%%%%%%%%%%%%%%%%%%%%%%%%%%%%%%%%%%%%%%



\bibliographystyle{mnras}
\bibliography{globulars} % if your bibtex file is called example.bib

%%%%%%%%%%%%%%%%%%%%%%%%%%%%%%%%%%%%%%%%%%%%%%%%%%

%%%%%%%%%%%%%%%%% APPENDICES %%%%%%%%%%%%%%%%%%%%%

%\appendix


%Additional material which would interrupt the flow of the main paper.

%%%%%%%%%%%%%%%%%%%%%%%%%%%%%%%%%%%%%%%%%%%%%%%%%%

% Don't change these lines
\bsp	% typesetting comment
\label{lastpage}
\end{document}

%% file: table_hst_obvs.tex
\begin{table*}
\centering
\caption{List of newly analysed \textit{HST} observations used for this project.}
\label{tabHSTobs}
\small{
\begin{tabular}{lllllc}
%\multicolumn{6}{c}{\textbf{Table~1.} List of \textit{HST} observations used for this project.}\\
\hline
Cluster & Camera & Filter & Exposures & Program ID & Epoch\\
\hline
NGC 5946 & WFC3/UVIS & F438W & $3\times500$\,s & 11628 & 2009.6\\
         &           & F555W & $3\times80$\,s  &       &       \\
NGC 6325 & WFC3/UVIS & F438W & $3\times435$\,s & 11628 & 2010.3\rule{0pt}{3ex}\\
         &           & F555W & $3\times85$\,s  &       &       \\
NGC 6342 & WFC3/UVIS & F438W & $3\times420$\,s & 11628 & 2009.6\rule{0pt}{3ex}\\
         &           & F555W & $3\times80$\,s  &       &       \\
NGC 6355 & WFC3/UVIS & F438W & $3\times440$\,s & 11628 & 2009.6\rule{0pt}{3ex}\\
         &           & F555W & $3\times80$\,s  &       &       \\
NGC 6380 & WFC3/UVIS & F555W & $3\times440$\,s & 11628 & 2010.2\rule{0pt}{3ex}\\
         &           & F814W & $3\times80$\,s  &       &       \\
NGC 6453 & WFC3/UVIS & F438W & $3\times450$\,s & 11628 & 2010.4\rule{0pt}{3ex}\\
         &           & F555W & $3\times80$\,s  &       &       \\
NGC 6517 & WFC3/UVIS & F555W & $3\times420$\,s & 11628 & 2010.3\rule{0pt}{3ex}\\
         &           & F814W & $3\times100$\,s &       &       \\
NGC 6553 & ACS/WFC   & F435W & $3\times340$\,s & 10753 & 2006.3\rule{0pt}{3ex}\\
         &           & F555W & $1\times 300$\,s &&\\
         &           & F814W & $1\times 60$\,s &&\\
\hline
\end{tabular}}
\end{table*}

%% file: table_gc_parameters.tex
\begin{table*}
\caption{Derived structural parameters for the 59 Milky Way globular clusters considered in this work. From left to right we list the GC name; metallicity \citep{1996AJ....112.1487H,2010arXiv1012.3224H}; mass and associated error; mass-to-light ratio and error; (1D) central velocity dispersion; total number of stars; and total number of WAGGS stars in the sample. Additional parameter values can be found at {\url{https://people.smp.uq.edu.au/HolgerBaumgardt/globular/}}}
\label{table_gc_params}
\begin{tabular}{lcccccccc}
\hline
GC Name & [Fe/H] & Mass & $\Delta$ Mass & M/L$_{\text{V}}$ & $\Delta$ M/L$_\text{V}$ & $\sigma_0$ & N$_{\text{total}}$ & N$_{\text{WAGGS}}$ \\
 & dex & \multicolumn{2}{c}{x $10^5$ M$_{\sun}$} & \multicolumn{2}{c}{M$_{\sun}$/L$_{\sun}$} & km/s & & \\
\hline
NGC 104 & -0.72 & 8.07 & 0.05 & 1.82 & 0.17 & 12.3 & 3254 & 78 \\
NGC 288 & -1.38 & 1.21 & 0.03 & 2.39 & 0.17 & 3.3 & 528 & 1 \\
NGC 362 & -1.26 & 3.37 & 0.05 & 1.7 & 0.25 & 8.9 & 479 & 19 \\
NGC 1261 & -1.27 & 1.73 & 0.15 & 2.2 & 0.38 & 5.6 & 288 & 39 \\
NGC 1851 & -1.18 & 2.83 & 0.04 & 1.92 & 0.14 & 10.6 & 669 & 58 \\
NGC 2298 & -1.92 & 0.54 & 0.1 & 2.11 & 0.42 & 3.5 & 40 & 8 \\
NGC 2808 & -1.14 & 8.18 & 0.06 & 1.67 & 0.12 & 15.0 & 1135 & 88 \\
NGC 3201 & -1.59 & 1.46 & 0.05 & 2.46 & 0.42 & 4.4 & 721 & 14 \\
NGC 4590 & -2.23 & 1.29 & 0.11 & 2.33 & 0.46 & 3.9 & 248 & 11 \\
NGC 4833 & -1.85 & 2.03 & 0.12 & 1.29 & 0.25 & 4.9 & 162 & 17 \\
NGC 5024 & -2.10 & 4.28 & 0.35 & 1.79 & 0.21 & 6.2 & 334 & 21 \\
NGC 5272 & -1.50 & 3.61 & 0.16 & 1.57 & 0.22 & 8.0 & 668 & 24 \\
NGC 5286 & -1.69 & 3.79 & 0.17 & 1.39 & 0.13 & 9.4 & 523 & 37 \\
NGC 5904 & -1.29 & 3.66 & 0.06 & 1.47 & 0.19 & 7.8 & 827 & 36 \\
NGC 5927 & -0.49 & 3.44 & 0.03 & 2.17 & 0.3 & 7.1 & 395 & 60 \\
NGC 5946 & -1.29 & 0.75 & 0.20 & 1.17 & 0.35 & 5.5 & 33 & 24 \\
NGC 5986 & -1.59 & 3.31 & 0.24 & 1.95 & 0.27 & 8.2 & 237 & 15 \\
NGC 6121 & -1.16 & 0.90 & 0.02 & 1.96 & 0.11 & 4.7 & 2817 & 11 \\
NGC 6171 & -1.02 & 0.77 & 0.04 & 2.19 & 0.61 & 4.1 & 373 & 22 \\
NGC 6218 & -1.37 & 0.82 & 0.04 & 1.51 & 0.34 & 5.0 & 495 & 13 \\
NGC 6254 & -1.56 & 1.88 & 0.04 & 1.8 & 0.06 & 6.2 & 406 & 16 \\
NGC 6304 & -0.45 & 1.61 & 0.14 & 2.12 & 0.38 & 5.3 & 171 & 5 \\
NGC 6325 & -1.25 & 0.73 & 0.12 & 1.65 & 0.45 & 6.3 & 42 & 34 \\
NGC 6333 & -1.77 & 3.16 & 0.24 & 2.28 & 0.22 & 8.2 & 34 & 25 \\
NGC 6342 & -0.55 & 0.60 & 0.11 & 3.64 & 1.52 & 5.6 & 49 & 33 \\
NGC 6352 & -0.64 & 0.55 & 0.02 & 1.79 & 0.17 & 3.3 & 40 & 8 \\
NGC 6355 & -1.37 & 0.71 & 0.14 & 0.57 & 0.15 & 5.2 & 50 & 32 \\
NGC 6356 & -0.40 & 3.82 & 0.80 & 1.55 & 0.36 & 7.6 & 44 & 23 \\
NGC 6362 & -0.99 & 1.08 & 0.03 & 2.25 & 0.12 & 3.6 & 342 & 3 \\
NGC 6380 & -0.75 & 3.05 & 0.02 & 2.06 & 2.73 & 8.1 & 46 & 36 \\
NGC 6388 & -0.55 & 10.4 & 0.09 & 1.87 & 0.23 & 17.7 & 511 & 26 \\
NGC 6397 & -2.02 & 0.89 & 0.01 & 2.23 & 0.35 & 5.4 & 2399 & 11 \\
NGC 6441 & -0.46 & 11.7 & 0.11 & 2.01 & 0.13 & 18.9 & 214 & 10 \\
NGC 6453 & -1.50 & 2.33 & 0.02 & 3.54 & 0.99 & 7.7 & 21 & 19 \\
NGC 6496 & -0.46 & 0.84 & 0.12 & 1.84 & 1.08 & 3.3 & 127 & 7 \\
NGC 6517 & -1.23 & 3.04 & 0.55 & 2.18 & 0.71 & 11.5 & 32 & 16 \\
NGC 6528 & -0.11 & 0.60 & 0.07 & 1.5 & 0.43 & 5.0 & 159 & 117 \\
NGC 6541 & -1.81 & 2.50 & 0.08 & 1.52 & 0.53 & 8.7 & 206 & 49 \\
NGC 6553 & -0.18 & 3.01 & 0.16 & 2.12 & 0.28 & 7.9 & 499 & 138 \\
NGC 6584 & -1.50 & 1.16 & 0.02 & 1.43 & 0.35 & 4.1 & 26 & 20 \\
NGC 6624 & -0.44 & 0.62 & 0.02 & 0.99 & 0.12 & 5.9 & 344 & 6 \\
NGC 6637 & -0.64 & 1.48 & 0.16 & 0.84 & 0.16 & 6.4 & 62 & 48 \\
NGC 6642 & -1.26 & 0.25 & 0.07 & 0.71 & 0.2 & 4.1 & 20 & 13 \\
NGC 6652 & -0.81 & 0.46 & 0.08 & 1.48 & 0.42 & 5.1 & 40 & 37 \\
NGC 6656 & -1.70 & 4.05 & 0.04 & 1.93 & 0.07 & 8.6 & 800 & 19 \\
NGC 6681 & -1.62 & 1.13 & 0.02 & 1.95 & 0.27 & 7.1 & 52 & 21 \\
NGC 6715 & -1.49 & 15.9 & 0.19 & 2.18 & 0.1 & 16.9 & 533 & 16 \\
NGC 6717 & -1.26 & 0.36 & 0.08 & 2.09 & 0.58 & 3.2 & 17 & 14 \\
NGC 6723 & -1.10 & 1.73 & 0.11 & 1.89 & 0.29 & 5.5 & 368 & 20 \\
NGC 6752 & -1.54 & 2.30 & 0.03 & 2.15 & 0.34 & 8.3 & 1184 & 37 \\
NGC 6760 & -0.40 & 2.57 & 0.30 & 2.01 & 0.27 & 7.2 & 80 & 53 \\
NGC 6809 & -1.94 & 1.87 & 0.07 & 2.79 & 0.55 & 4.9 & 492 & 18 \\
NGC 6838 & -0.78 & 0.53 & 0.03 & 2.8 & 1.04 & 2.9 & 256 & 14 \\
NGC 6934 & -1.47 & 1.40 & 0.25 & 1.73 & 0.31 & 4.9 & 45 & 12 \\
NGC 6981 & -1.42 & 0.68 & 0.12 & 1.22 & 0.23 & 3.0 & 21 & 19 \\
NGC 7006 & -1.52 & 1.47 & 0.38 & 1.94 & 0.52 & 4.0 & 43 & 18 \\
NGC 7078 & -2.37 & 4.94 & 0.05 & 1.18 & 0.11 & 13.6 & 1273 & 6 \\
NGC 7089 & -1.65 & 5.05 & 0.10 & 1.68 & 0.06 & 10.7 & 410 & 21 \\
NGC 7099 & -2.27 & 1.27 & 0.07 & 1.8 & 0.38 & 5.5 & 709 & 6 \\
\hline
\end{tabular}
\end{table*}